\documentclass[graybox, envcountchap]{svmult}

\usepackage{mathptmx}        
\usepackage{amsmath}
\usepackage{mathtools}
\usepackage{amssymb}
\usepackage{color}
\usepackage{helvet}          
\usepackage{courier}         
\usepackage{dirtree}

\usepackage{makeidx}        
\usepackage{graphicx}        
\usepackage{subfig}

\usepackage{multicol}        
\usepackage[bottom]{footmisc}

\usepackage{hyperref}        
\hypersetup{colorlinks=true,urlcolor=blue}

\usepackage[misc]{ifsym}

\usepackage[numbers]{natbib}
\newcommand{\edit}[1]{{#1}}

\newcommand{\sgrastar}[0]{Sgr A$^{*}$}

\usepackage{acronym}

\newacro{BH}{black hole}
\newacro{BHL}{Bondi-Hoyle-Lyttleton}
\newacro{BZ}{Blandford-Znajek}
\newacro{ECO}{\textit{exotic compact object}}
\acrodefplural{ECO}[ECOs]{\textit{exotic compact objects}}
\newacro{EHT}{Event Horizon Telescope}
\newacro{GMGHS}{Gibbons-Maeda-Garfinkle-Horowitz-Strominger}
\newacro{GRHD}{general relativistic hydrodynamics}
\newacro{GR}{general relativity}
\newacro{GRMHD}{general relativistic magnetohydrodynamics}
\newacro{GRRT}{general relativistic radiative transfer}
\newacro{HBH}{hairy black hole}
\newacro{ISCO}{innermost stable circular orbit}
\newacro{MAD}{magnetically arrested disk}
\newacro{MHD}{magnetohydrodynamics}
\newacro{MRI}{magneto-rotational instability}
\newacro{QPO}{quasi-periodic oscillation}
\newacro{SANE}{standard and normal evolution}
\newacro{SMBH}{supermassive black hole}
\newacro{VLBI}{very long baseline interferometry}

\makeindex             

\begin{document}


\title{GRMHD simulations of accretion onto exotic compact objects}
\author{Héctor R. Olivares Sánchez, Prashant Kocherlakota,
Carlos A. R. Herdeiro}
\institute{Héctor R. Olivares Sánchez (\Letter) \at Departamento de Matemática da Universidade de Aveiro and Centre for Research and Development in Mathematics and Applications (CIDMA), Campus de Santiago, 3810-193 Aveiro, Portugal, \email{h.sanchez@ua.pt}
\and Prashant Kocherlakota \at 
Black Hole Initiative at Harvard University, 20 Garden St., Cambridge, MA 02138, USA \\
Center for Astrophysics, Harvard \& Smithsonian, 60 Garden St., Cambridge, MA 02138, USA\\ 
\email{pkocherlakota@fas.harvard.edu}
\and Carlos A. R. Herdeiro \at Departamento de Matemática da Universidade de Aveiro and Centre for Research and Development in Mathematics and Applications (CIDMA), Campus de Santiago, 3810-193 Aveiro, Portugal, \email{herdeiro@ua.pt}}
%
%
\maketitle

\abstract{Some of the extensions to general relativity and to the Standard Model of particle physics predict families of hypothetical compact objects, collectively known as exotic compact objects (ECOs). This category can be defined to encompass non-Kerr black holes both within and beyond general relativity, as well as horizonless compact objects such as boson stars. In order to model observational signatures and identify possible detections, it is crucial to understand the interaction between these objects and their surrounding medium, usually plasmas described by the equations of general relativistic magnetohydrodynamics (GRMHD). To this end, we review the existent literature on GRMHD simulations of accretion onto these objects. These cover a variety of objects and accretion patterns. We conclude by listing possible directions to continue exploring this relatively young field.}


\section{Introduction}
\label{sec:intro}
Many of the effects from
extensions to \ac{GR} and the Standard Model of particle physics are expected to manifest in strong gravity environments, and are directly linked to relativistically
compact objects.
In fact, some of them directly predict new families of
hypothetical compact objects, collectively known
as \acp{ECO}.
The advances in observing technologies of the last decades, such as Very Long Baseline Interferometry (VLBI), gravitational wave
(GW) detectors, and the possibility of multi-messenger astronomy, 
allow us to zoom into the environments of compact objects,
with a great potential for new discoveries in astro- and fundamental physics.
However, without a proper understanding of the interaction
between these objects and their environments,
 we risk to misinterpret environmental
effects as new fundamental physics, or to miss the opportunity to make important discoveries.

Most of the baryonic matter in the Universe is
in the state of plasma, and in general can be
well described by the equations of \ac{GRMHD}.
Understanding the observational properties of
\acp{ECO} in realistic astrophysical environments
is one of the motivations to perform \ac{GRMHD}
simulations of accretion onto these objects.

Although the field is still quite young, in this review
we aim to encompass the existent literature
featuring such simulations. We will include not only
\ac{GRMHD} simulations, but also
simulations performed in pure \ac{GRHD}.

This chapter is organized as follows.
In Section \ref{sec:landscape-of-ECOs},
we give an overview of some of the most interesting
families of \acp{ECO}, as well as some
pre-requisites for their consideration as viable models
for compact objects in our Universe.
Later, in Section \ref{sec:related-topics},
we discuss some important issues that is necessary
to consider from the point of view of modeling
before embarking in performing \ac{GRMHD} simulations.
An actual review of the results currently
available in the literature is presented
in Section \ref{sec:gmrhd-simulations}.
Finally, in Section \ref{sec:conclusion}
we conclude by giving some future perspectives
on the field.


\section{A view on the landscape of Exotic Compact Objects}
\label{sec:landscape-of-ECOs}
The most compact objects in the Universe are now widely believed to be \acp{BH}. Under the \textit{Kerr hypothesis}, moreover, they are believed to be well modelled by the vacuum \acp{BH} of \ac{GR}~\cite{Schwarzschild:1916uq,Kerr:1963ud}. For a large part of their century-old history, however, \acp{BH} were speculative, or exotic, solutions of \ac{GR}. Observational evidence~\cite{Narayan2013a} together with the (unexpected) theoretical robustness of these solutions~\cite{Regge:1957td,Zerilli:1970se,Whiting:1988vc}, slowly established their case as physical objects, abundantly realized in the Cosmos. Yet, their subtle phenomenology together with the theoretical puzzles they introduce~\cite{Penrose:1964wq,Hawking:1976ra,Almheiri:2012rt} invite us 
 to ask if the astrophysical \ac{BH} \textit{candidates} are - all of them~\cite{Herdeiro:2022yle} - well modelled by the Kerr solution.

 There are two qualitatively different departures from the Kerr hypothesis. The first way is that the astrophysical \acp{BH}s candidates are in fact \acp{BH}, in the sense of \ac{GR}, but not Kerr \acp{BH}. It is important to observe they could be either non-Kerr \acp{BH} in \ac{GR} or beyond \ac{GR}, as will be expanded below. The second way is that the astrophysical BH candidates are not \acp{BH}; they are compact objects without a horizon. We shall call all such speculative, alternative, dark compact objects  \acp{ECO}, encompassing both \textit{non-Kerr \acp{BH}} and \textit{horizonless ECOs}.

\subsection{Theoretical robustness}
\label{sec:theoretical-robustness}
Precise strong gravity observables, such as gravitational waves and \ac{BH} imaging, are largely theory-driven: one needs theoretical templates in order to interpret the results. By the same token, one needs such templates for \acp{ECO}, if these are to be considered as contenders for the data and to allow a hypothetical detection. 

Still, before embarking on the considerable task of obtaining precision phenomenological templates for a generic \ac{ECO} model, the model should have surpassed several important theoretical criteria, making it worthy of the phenomenological endeavour. 

The first criterion is that the \ac{ECO} should arise in a consistent theoretical model. This does not requires the model to be complete. \ac{GR}, for instance, is not UV complete; but within its regime of validity it is self-consistent, in particular offering a well-posed Cauchy problem (the hyperbolicity issue)~\cite{Choquet-Bruhat:1969ywq}. This is by no means guaranteed for an \ac{ECO} model. It is elucidative to consider two examples of its failure. The first one is in \ac{GR} with ``matter". Consider the ``matter" to be a Proca field, which is appealing because it leads to a popular model of horizonless \acp{ECO} called Proca stars~\cite{Brito:2015pxa}. Including apparently innocuous self-interactions for the Proca field leads to a breakdown of hyperbolicity~\cite{Coates:2022nif,Coates:2022qia,Clough:2022ygm,Mou:2022hqb}. The second one is in modified gravity. Consider Einstein-scalar-Gauss-Bonnet models, which are appealing within higher curvature models because the equations of motion are still second order. Additionally, it has known examples of interesting non-Kerr \acp{BH}~\cite{Kanti:1995vq,Kleihaus:2011tg,Silva:2017uqg,Doneva:2017bvd,Antoniou:2017acq,Cunha:2019dwb,Herdeiro:2020wei,Berti:2020kgk}. However, it was observed that there is a breakdown of hyperbolicity in dynamical evolutions, at least for some classes of scalar-curvature couplings~\cite{Ripley:2019aqj}.

The second criterion is that the \ac{ECO} should be sufficiently stable, so that it could play a role in astrophysical processes. Establishing mode stability of a given \ac{ECO} model may be rather challenging, especially beyond spherical symmetry. Moreover there may be even more subtle instabilities, possibly unseen in a mode stability analysis. One such pathology of recent interest in the context of horizonless \acp{ECO} is the \textit{light ring instability}, or \textit{trapping instability}. Light rings are circular, bound, photon orbits that generically exist around \acp{BH}~\cite{Cunha:2020azh}. They are closely connected to strong gravity phenomenology, such as the ringdown of a perturbed \ac{BH}/\ac{ECO}~\cite{Cardoso:2016rao} and the \ac{BH} shadow~\cite{Cunha:2017eoe}. For horizonless \acp{ECO} that can emerge from an incomplete gravitational collapse in \ac{GR}, when they become sufficiently compact to have light rings, these come in pairs~\cite{Cunha:2017qtt}. Together with a light ring similar to that of Kerr \acp{BH}, that is associated to the aforementioned strong gravity phenomenology, there is a second light ring which, unless, the null energy condition is violated, is a stable, bound photon orbit. These stable orbits are potentially problematic; they form a potential well that can trap modes of null fields (such as gravity), leading to a strong backreaction and thus a spacetime instability. It was argued that this instability is non-linear~\cite{Keir:2014oka}; moreover, the instability was confirmed to have short time scales in concrete \ac{ECO} models~\cite{Cunha:2022gde}. As such, this may be a generic issue with horizonless \ac{ECO} models that have light rings and have a clear formation mechanism. Another possible cause of instability of rotating horizonless \acp{ECO} is the existence of an ergoregion~\cite{1978RSPSA.364..211C}. It has been shown, however, that generically the existence of an ergoregion in horizonless \acp{ECO} is accompanied by light rings~\cite{Ghosh:2021txu}; in other words, light rings arise first when making a spinning object compact, making their associated instability a more fundamental one. 

A third and final criterion is that the \ac{ECO} should have a formation mechanism. Non-Kerr \acp{BH} in \ac{GR} or modified gravity may, in general, form by gravitational collapse, in appropriate circumstances. But the formation of horizonless \acp{ECO} may be more delicate. Popular models, such as gravastars~\cite{Mazur:2004fk}, fuzzballs~\cite{Mathur:2008nj} or wormholes~\cite{Morris:1988cz}, have no clear, under control formation mechanism, that has been dynamically tested, even though some general principles may exist. By contrast, a model of horizonless \acp{ECO} with a tested formation mechanism is the family of bosonic stars, the mechanism being known as gravitational cooling~\cite{Seidel:1993zk}. Together with the established stability of some of these stars (in regions of  their parameter space), this has made bosonic stars a popular model of horizonless \acp{ECO}.

\subsection{Horizonless ECOs}
\label{sec:horizonless-ECOs-theory}
Many models of horizonless \acp{ECO} have been proposed throughout the years - see the review~\cite{Cardoso:2019rvt}. They have had different motivations, not always to replace \acp{BH} and solve their aforementioned theoretical puzzles; some, in fact, co-exist with \acp{BH}. The latter have other motivations, that could be theoretical - such as geons~\cite{Wheeler:1955zz}, to model classical particles in field theory - or phenomenological - such as bosonic stars (see the review~\cite{Liebling:2012fv}) - that could relate to fuzzy dark matter~\cite{Hui:2016ltb}. Most of these, however, do not fulfil one or more of the above criteria. Whereas this does not prevent some phenomenological considerations, e.g. computing geodesics or imaging properties under some academic models, or even full \ac{GRMHD} simulations on a fix background, their linear and (especially) non-linear dynamics is often still out of reach. 

One exception in which dynamics is under control is the broad family of bosonic stars. These are lumps of a complex, massive scalar or vector field. The scalar stars, known as boson stars, have been known for over half a century~\cite{Kaup:1968zz,Ruffini:1969qy}, although their rotating versions were only found much later~\cite{Schunck:1996he,Yoshida:1997qf}. The vector stars, known as Proca stars, have only been constructed and studied over the last decade~\cite{Brito:2015pxa} - see also~\cite{Herdeiro:2019mbz}. To be in the astrophysical \acp{BH} mass range, these stars - at least in the simplest models - must be composed by ultralight bosonic particles, in fact, as a Bose-Einstein condensate of very many such ultralight particles in the same quantum state. Their collapse into a \ac{BH} is prevented from Heisenberg's uncertainty principle, up to a (model dependent) maximal mass. They can form via gravitational cooling~\cite{Seidel:1993zk} and many models (static and spinning) have dynamically stable regions in the parameter space~\cite{Liebling:2012fv}, even though some unexpected instabilities have been found as well~\cite{Sanchis-Gual:2019ljs,Herdeiro:2023wqf}.

It is not surprising, therefore, that there is a considerable literature on strong gravity phenomenology of bosonic stars (by which we mean both the scalar and vector cases, collectively). In particular \ac{GRMHD} simulations on the background of bosonic stars have been performed, as described  below~\cite{olivares_how_2020}. Since some of the models in~\cite{olivares_how_2020} are  not dynamically stable, this has led to exploring the imaging of other bosonic stars, within the program of the \textit{imitation game}: assessing possible degeneracies in imaging properties between \acp{BH} and horizonless \acp{ECO}. In~\cite{herdeiro_imitation_2021} it was observed that spherical Proca stars could produce an effective shadow, due to a special property of timelike circular orbits, that could quench the \ac{MRI} driving matter inwards in an accretion disk, effectively creating a void of emitting matter near the centre of the star. It was even possible to find a Proca star for which this effective shadow is similar in size to that of a Schwarzschild \ac{BH} with the same mass. Subsequently, however, it was found that such Proca star have an unexpected instability against non-spherical dynamics~\cite{Herdeiro:2023wqf}, undermining this model of a \ac{BH} foil. Still, as a proof of principle, it would be interesting to understand if such void creating mechanism actually holds in a more complete \ac{GRMHD} analysis. 

\begin{figure}
	\centering
	\begin{tabular}{cc}
		\subfloat[$\theta=5^o$ ]{\includegraphics[width=0.5\columnwidth]{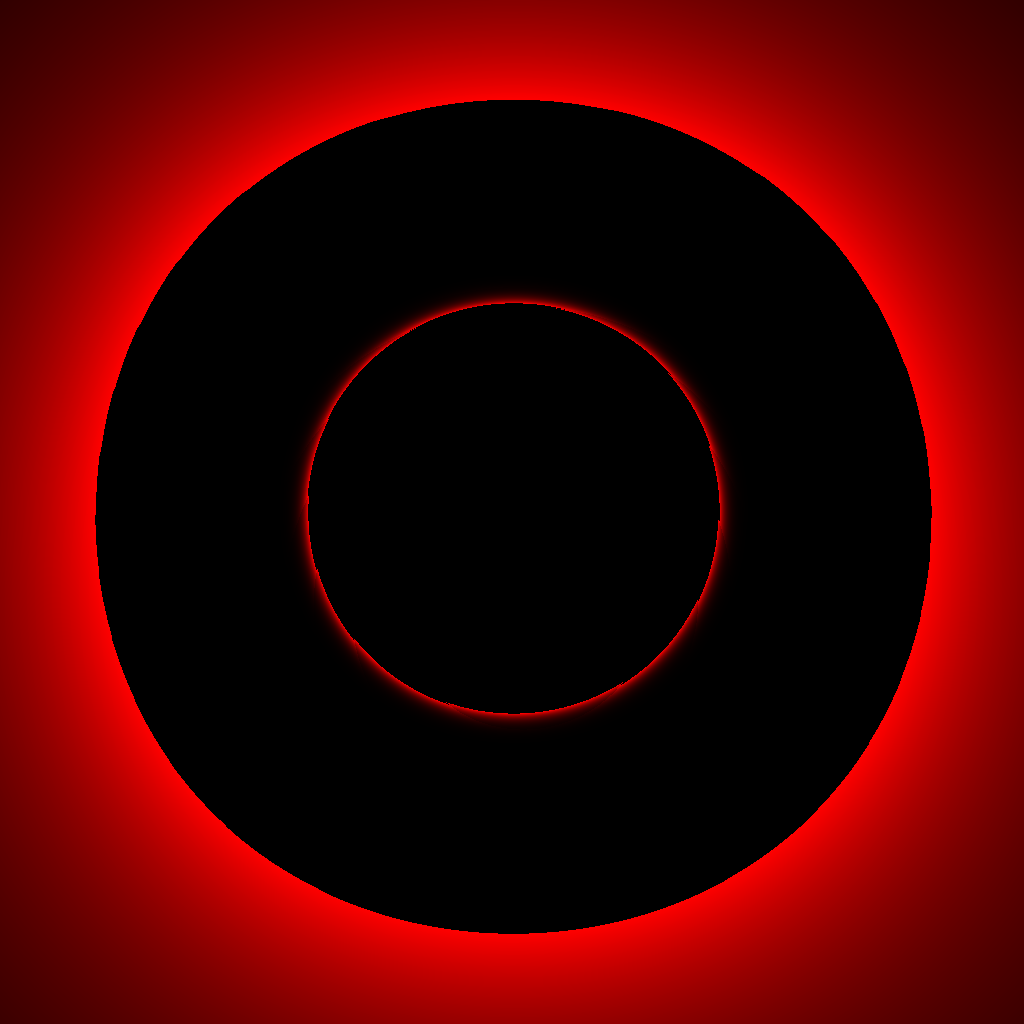}} 
		\subfloat[$\theta=88^o$ ]{\includegraphics[width=0.5\columnwidth]{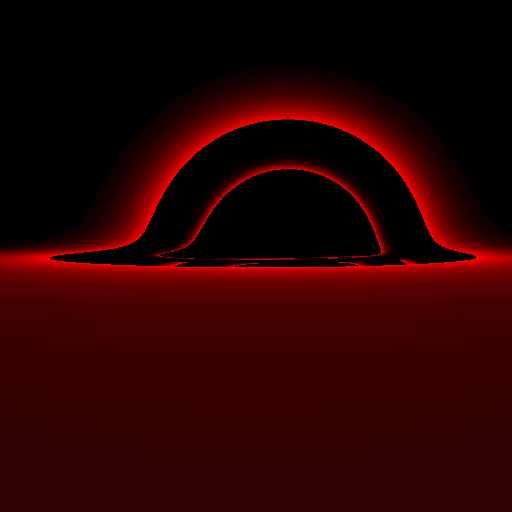}}\\
	\end{tabular}
	\caption{\small \label{fig:imitation} Lensing images for a spinning Proca star without light rings but with discontinuous regions of equatorial, timelike circular orbits, wherein emitting matter is placed. The images are for two different co-latitude $\theta$ observations. From \cite{sengo_imitation_2024}.}
\end{figure}

Spinning Proca stars, on the other hand, are dynamically robust in their fundamental state~\cite{Sanchis-Gual:2019ljs}. It was observed that these stars have a peculiar structure of their timelike, equatorial, counter-rotating, circular orbits, with disconnected regions of stable orbits~\cite{Delgado:2021jxd}. This has led to the suggetions that matter accumulated in the regions of such stable orbits could mimic some imaging features of \acp{BH}~\cite{sengo_imitation_2024}: not only a shadow, but also fine features attributed to the existence of light rings, but in a spacetime without light rings - Fig.~\ref{fig:imitation}. Again, it would be interesting to test if such features are kept or washed way in a more realistic \ac{GRMHD} analysis. For other imaging studies of bosonic stars and their comparison with \acp{BH} see e.g.~\cite{Vincent:2015xta,Cunha:2015yba,Cunha:2016bjh,Rosa:2022tfv,Rosa:2022toh,Rosa:2023qcv,Rosa:2024eva}.

\subsection{Non-Kerr \acp{BH}}
\label{sec:non-Kerr-BHs-theory}
The spectrum of non-Kerr \acp{BH} both in \ac{GR} and beyond \ac{GR} is immense. Imaging properties of such \acp{ECO} are now a very popular subject with hundreds of models analysed in more or less academic setups. Again, however, restricting to \acp{ECO} abiding the above criteria narrows down significantly the space of models. But since most of imaging analyses rely simply on computing geodesics, this program can be pursued even if the model lacks a strong theoretical support. A broad conclusion is that for most models current observations of Sgr A* and M87* can only constraint part (if any at all) of the parameter space.

\ac{GRMHD} simulations, on the other hand, have been considered only for a handful of models. One case is dilatonic \acp{BH}~\cite{Mizuno2018,Roder:2023oqa}, which is a \ac{GR} model,  since the dilatonic (and axionic) field are minimally coupled to gravity, while non-minimally coupled to the electromagnetic field. These explorations support the expectation that in real-world observations imaging approaches will not be able to clearly distinguish such alternative models from Kerr using the current finite resolution of the EHT. 

A similar conclusion can be found for many beyond \ac{GR} models: the non-Kerrness is small and differences hard to tell from imaging e.g.~\cite{Cunha:2016wzk,Cunha:2019dwb,Fernandes:2024ztk}. But examples where dramatic differences from Kerr occur are also known. One such family is that of \acp{BH} with synchronised hair, which interpolates between Kerr \acp{BH} and spinning bosonic stars~\cite{Herdeiro:2014goa,Herdeiro:2016tmi}. Some dramatic imaging features can occur, such as chaotic lensing and multiple shadows of a single object~\cite{Cunha:2015yba}. In part of the parameter space such solutions have a natural formation mechanism via the phenomenon of superradiance~\cite{Brito:2015oca}, but in this region deviations from Kerr are smaller~\cite{Cunha:2019ikd}. It would be quite interesting to perform \ac{GRMHD} simulations on these backgrounds. In this respect, equilibrium configurations of self-gravitating accretion disks have been computed, which could be taken as initial data for such simulations~\cite{Gimeno-Soler:2018pjd,Gimeno-Soler:2021ifv}. 

Let us comment that lensing has also been considered for even more exotic (and problematic) spacetimes, such as naked singularities~\cite{Ortiz:2015rma,Joshi:2020tlq,Wang:2023jop}. Pushing \ac{GRMHD} simulations on these background \edit{requires to face} issues related to the arbitrariness of boundary conditions near the singularity, even though this may be mitigated for repulsive singularities.
\edit{An interesting work performing \ac{GRHD} simulations on a spacetime containing naked singularities can be found in Reference \cite{bambi_accretion_2009}.}

\section{Practical considerations}
\label{sec:related-topics}

In broad terms, the aim of a \ac{GRMHD} simulation
can be
described as to increase the level of realism
of a model for a relativistically compact object
by modeling also its interactions with a surrounding
medium. In most of the works based on \ac{GRMHD}
simulations of \acp{ECO}, the simulation data
is further postprocessed to connect this modeling with
observables.

During this process, physical models are transformed into
numerical models, and it is crucial to ensure
that, during this process, the essential features of
physical models are preserved.
This translates into making some choices for the numerical
model, that are generally related to accommodating
the physical model to the limitations of a numerical model.
Here we briefly discuss a selection
of these practical considerations, and some of the choices
that require special attention when simulating
\acp{ECO} surrounded by the interstellar medium.

{\it General assumptions.}
All of the works reviewed in this chapter
make the assumption that the fluid that makes the interstellar
medium is non self-gravitating and its evolution does not
back-react on the spacetime metric,
so that the only source
of gravity is the \ac{ECO}. This is known as the
{\it test fluid approximation}, and is justified
for systems such X-ray binaries and \ac{SMBH} candidates,
where the central object is several orders of magnitude
more massive than the surrounding medium.
In addition to this, the spacetime metric is not
expected to evolve, so simulations are performed
on {\it stationary spacetimes}.
Other important assumptions are that the fluid
is described by a gas with a simple equation of
state (be it ideal or polytropic),
and (for those involving magnetic fields), that it is
a very good electric conductor. The latter is known
as the {\it ideal magnetohydrodynamics} assumption,
and it is generally an accurate description of interstellar
plasmas. Finally, even though some of the \acp{ECO}
that are simulated involve the presence of
additional fields (general scalar, axion or dilaton fields),
their couplings to the matter in the fluid
or to the simulated electromagnetic field are ignored.
Further comments on this assumption
can be encountered in parts of Section 
\ref{sec:gmrhd-simulations} dealing
with specific works in the literature .

{\it Flexibility to adopt custom metrics.}
To simulate most of the \ac{ECO} models described
in Section \ref{sec:landscape-of-ECOs},
it is essential to endow \ac{GRMHD} codes with the
capability of performing covariant simulations
on non-Kerr spacetimes.
For analytic metrics, this just involves replacing the
expressions for metric functions in the code by the
appropriate ones. The applicability of this
procedure can be significantly enlarged by the use of
parameterized expansions, that can represent
several theories by different values of the parameters
\cite{Mizuno2018,Nampalliwar2022,Chatterjee+2023a,Chatterjee+2023b}.
For cases where the metric is numerical
(an important example are bosonic stars and \acp{BH}
with synchronized hair),
simulations may resort to tabulated metrics
\cite{Olivares2019a,teodoro_tidal_2021} or to evaluations
of spectral representations of the metric
\cite{Meliani2016,meliani_tidal_2017}.

{\it Initial conditions.}
Due to the finiteness in time of numerical models,
it is necessary to specify the configuration
of the fluid at the start of the simulation.
Naturally, this is problem-dependent,
but we can mention here some generalities.
Most of the simulations of magnetized accretion
onto \acp{BH} and other compact objects start from
tori in hydrodynamic equilibrium to which magnetic
loops are added.
Even though these are the most widely used initial
condition for this type of simulations,
their applicability is limited to initially small
magnetic fields, as their inclusion breaks this
equilibrium \cite{cruz-osorio_non-linear_2020}.
Solutions for magnetized tori in true equilibrium have
been obtained for boson stars and \acp{BH} with scalar hair
in \cite{Gimeno-Soler:2018pjd,Gimeno-Soler:2021ifv},
although they have so far not been utilized in simulations.
In fact, the sub-field of equilibrium solutions
in this kind of spacetimes features quite interesting
results, that differ in significant ways from those
that are possible around Kerr \acp{BH}
(see e.g.
\cite{teodoro_retrograde_2021,teodoro_scalarizedBHs_2021}).

{\it Boundary conditions.}
We also need to specify the behavior
of the fluid when interacting with the spatial
boundaries of the simulation. Simulations
of \acp{ECO} can present very different
challenges to those of accretion onto Kerr \acp{BH}.
For objects that possess an event horizon,
it is customary to extend the simulation domain
slightly inside it in order to prevent
signals associated to the boundary
from propagating into the simulation
and producing spurious effects
(see e.g. \cite{cruz-osorio_flip-flop_2012}).
For this reason, it is important to use
horizon penetrating expressions for the metric.
These have been obtained for some general classes
of metrics, and due to the importance of the
subject, we have dedicated Section
 \ref{sec:horizon-penetrating-coords}
to the procedure for obtaining them.

For the case of horizonless compact objects,
there are two possibilities: either the simulation
domain has a boundary at a physical surface,
or it continues down to the center of the object.
\acp{ECO} with surfaces have not been considered
so far for \ac{GRMHD} simulations,
however there are works in the literature
that simulate accretion onto objects with
physical surfaces such as neutron stars
(see Section \ref{sec:horizonless-ECOs-grmhd}).
For cases where the center of the object is part
of the computational domain, care must be taken
to avoid problems related to the coordinate singularity
at the origin of spherical coordinates.
This can be done by selecting boundary conditions
with the appropriate symmetry properties or by ensuring
communication between neighboring cells at each
side of the singularity \cite{Meliani2016,olivares_how_2020}.
Another approach is to remove the coordinate
singularity by resorting to Cartesian coordinate systems
(see e.g. \cite{meliani_tidal_2017,teodoro_tidal_2021}).
\edit{
The case of \ac{GRHD} simulations in spacetimes
with true curvature singularities
was explored in \cite{bambi_accretion_2009}.
Although the proper way to treat numerically domains that
contain such singularities is still an open problem;
one can consider (as argued by the authors of \cite{bambi_accretion_2009})
that a naked singularity is a pathology of classical \ac{GR} which should be replaced
by something else in a yet unknown full theory.
This motivates setting, for instance, absorbing
boundary conditions around the region containing the unknown physics,
and excising it from the domain, as done in the cited work.}


\subsection{Horizon-Penetrating Coordinates}
\label{sec:horizon-penetrating-coords}
Most current codes used to perform GMRHD simulations require, reasonably, the spacetime metric in a horizon-penetrating form, to avoid numerical issues due to the presence of the coordinate singularity at the horizon. For the Kerr metric, specific coordinates, suitable for simulations, were introduced in Refs. \cite{Font+1998, McKinney+2004}. A large class of geodesically-integrable, asymptotically-flat \ac{BH} spacetimes, characterized by four free functions, were discovered in Ref. \cite{Johannsen2013}. The horizon-penetrating Kerr-Schild form of such Johannsen(-Psaltis) spacetimes was also obtained there (Sec. V) and has recently been used to simulate accretion in such spacetimes \cite{Nampalliwar2022, Chatterjee+2023b}.

The horizon-penetrating form for yet another broad class of stationary and axisymmetric spacetimes was reported in Ref. \cite{Kocherlakota+2023}. The Azreg-A{\"i}nou (AA) metric, used there, was discovered in Refs. \cite{Azreg-Ainou2014a, Azreg-Ainou2014b, Azreg-Ainou2014c} in an attempt to obtain stationary and axisymmetric solutions, through an algorithmic procedure similar to the one employed by Newman \& Janis \cite{Newman+1965}, from a static and spherically-symmetric ``seed'' metric. The AA metric has the value of describing null geodesically-integrable spacetimes, rigidly rotating \acp{BH}, and of capturing several well-known solutions to \ac{GR} and alternative theories, some of which we discuss in Sec. \ref{sec:Non-Kerr-BHs-GRMHD} below.

We now briefly review the steps involved in arriving at coordinates that are suitable for \ac{GRMHD} simulations, using the AA metric as a simple case study. Its line element in Boyer-Lindquist (BL; \cite{Boyer+1967}) coordinates, $x^\mu = (t, r, \vartheta, \varphi)$, is given as \cite{Azreg-Ainou2014c, Kocherlakota+2023}
\begin{align} \label{eq:AA_Metric_BL}
\mathrm{d}s^2 =&\ \frac{X}{\Sigma}\left[-\left(1-\frac{2F}{\Sigma}\right)\mathrm{d}t^2 - 2\left(\frac{2F}{\Sigma}\right)a\sin^2{\vartheta}\mathrm{d}t\mathrm{d}\varphi + \frac{\Pi}{\Sigma}\sin^2{\vartheta}\mathrm{d}\varphi^2 \right. \\
&\ \left. + \frac{\Sigma}{\Delta}\mathrm{d}r^2 + \Sigma\mathrm{d}\vartheta^2\right]\,, \nonumber
\end{align}
where $a$ denotes the spin parameter and the metric functions are not all independent. In addition to the conformal factor $X$, there are only two other free functions, $A=A(r)$ and $B=B(r)$. The latter pair determines the remaining metric functions via $2F=A-B,\ \Delta=B+a^2,\ \Sigma=A+a^2\cos^2{\vartheta},$ and $\Pi=(A+a^2)^2 - \Delta a^2\sin^2{\vartheta}$. 

To obtain the Kerr metric, one sets $X=\Sigma$, $A=r^2$, and $B=r^2 - 2Mr$. For the Kerr-Newman metric \cite{Newman+1965b}, which is the stationary electrovacuum \ac{BH} solution of the Einstein-Maxwell equations, $B$ is modified to $B=r^2 - 2Mr + Q^2$. In the above, $M$ and $Q$ are the Arnowitt-Deser-Misner (ADM; \cite{Arnowitt+2008}) mass and charge of the spacetime respectively.

When used to describe \ac{BH} spacetimes, horizons are permitted at the roots of $\Delta$, where the $rr-$component of the metric diverges. This coordinate singularity can be eliminated by moving to coordinates, $x^{\bar{\mu}} = (\tau, r, \vartheta, \phi)$, in which the one-form field associated with the ingoing principal null congruence (PNC) of the spacetime, $(\ell_-)_\mu = \left[-\Delta, -\Sigma, 0, \Delta a\sin^2{\vartheta}\right]$, becomes $(\ell_-)_{\bar{\mu}} = [-1, -1, 0, a\sin^2{\vartheta}]$. The Jacobian, $\Lambda^{\bar{\mu}}_{\ \ \mu} = \partial_\mu x^{\bar{\mu}}$, for the desired transformation from BL to these so-called ``spherical ingoing Kerr-Schild'' (siKS) coordinates is given as
\begin{equation} \label{eq:BL_siKS_Jacobian}
\Lambda^{\bar{\mu}}_{\ \mu} = 
\begin{bmatrix}
1 & 2F/\Delta & 0 & 0 \\
0 & 1 & 0 & 0\\
0 & 0 & 1 & 0\\
0 & a/\Delta & 0 & 0
\end{bmatrix}\,.
\end{equation}
The horizon-penetrating form of the metric \eqref{eq:AA_Metric_BL} is then \cite{Kocherlakota+2023}
\begin{align} \label{eq:AA_Metric_siKS}
\mathrm{d}s^2 =&\ \frac{X}{\Sigma}\left[-\left(1-\frac{2F}{\Sigma}\right)\mathrm{d}\tau^2 + \left(1+\frac{2F}{\Sigma}\right)\mathrm{d}r^2 + \Sigma\mathrm{d}\vartheta^2 + \frac{\Pi}{\Sigma}\sin^2{\vartheta}\mathrm{d}\phi^2\right. \\
&\ \left. - 2\left(\frac{2F}{\Sigma}\right)a\sin^2{\vartheta}\mathrm{d}\tau\mathrm{d}\phi + 2\left(\frac{2F}{\Sigma}\right)\mathrm{d}\tau\mathrm{d}r - 2\left(1+\frac{2F}{\Sigma}\right)a\sin^2{\vartheta}\mathrm{d}r\mathrm{d}\phi\right]\,, \nonumber
\end{align}
from which we see that there are no divergences at the horizons ($\Delta=0$).

The two key ingredients therefore are finding both the 1-form field associated with the ingoing PNC in BL coordinates and the coordinate transformation that transforms it to an accepted standard form, $(\ell_-)_{\bar{\mu}} = [-1, -1, 0, a\sin^2{\vartheta}]$. Thus, the siKS coordinates are adapted to the ingoing PNC in a specific way, and, thus, penetrate the future horizon. The metric above has recently been used to perform \ac{GRMHD} simulations in ``dilaton-axion'' \ac{BH} spacetimes, as discussed below in Sec. \ref{sec:axion-dilaton-GRMHD}. Furthermore, the Jacobian \eqref{eq:BL_siKS_Jacobian} can be used to conveniently transform \ac{GRMHD} data back to BL coordinates, if of interest. 

The $3\!+\!1$ split of the metric \eqref{eq:AA_Metric_siKS} has also been discussed in Ref. \cite{Kocherlakota+2023}, from which it becomes clear that the fiducial choice for the congruence of hypersurface-forming Eulerian observers corresponds to a timelike one that has the same angular velocity as that of the zero angular momentum observer congruence.

It is worth noting that the closely-related set of spherical ingoing \textit{null} Kerr-Schild coordinates, $x^{\tilde{\mu}} = (v, r, \vartheta, \phi)$, more commonly used in the \ac{GR} literature, are obtained by transforming the ingoing PNC generator to the form $(\ell_-)_{\bar{\tilde{\mu}}} = [-1, 0, 0, a\sin^2{\vartheta}]$. From $\ell_-^{\bar{\tilde{\mu}}} = -\delta^{\bar{\tilde{\mu}}}_r$, we see that these are truly adapted to the ingoing PNC. For the Kerr metric, in particular, this form can be seen, e.g., in Sec. 5.3.6 of Ref. \cite{Poisson2004}. The Jacobian for the coordinate transformation from BL coordinates is similar to the one in eq. \ref{eq:BL_siKS_Jacobian} with the term $2F/\Delta$ being modified to $2F/\Delta + 1$.


\section{GRMHD simulations of accretion onto exotic compact objects}
\label{sec:gmrhd-simulations}

\subsection{Horizonless compact objects}
\label{sec:horizonless-ECOs-grmhd}
Accretion onto compact objects with no event horizon
is of particular interest, as it is expected to produce a qualitatively different phenomenology from that onto \acp{BH}, as matter does not disappear behind
the event horizon.
For compact objects with a surface, such as neutron
stars, one would expect an interaction between this surface and the accreting matter, which should
result in heating the surface.
However, in practice, \ac{GRMHD} simulations
of accretion onto neutron stars employ absorbing
boundary conditions \cite{parfrey_general-relativistic_2017} or simply remove the matter that
falls onto the surface from the simulation
\cite{das_three-dimensional_2023}.
Instead, the physics of the surface is usually accounted
for in post-processing, at the stage of calculating
images and spectra. For instance, 
\citep{EHT_SgrA_PaperV}
considers the cases of thermally emitting and reflecting
surfaces. The results of such analysis apply
also to \acp{ECO} with surfaces for which the
exterior metric is very close to that of Schwarzschild
and Kerr \acp{BH}, such as gravastars.

More interesting from the point of view of \ac{GRMHD} simulations are horizonless compact objects that also
lack surfaces.
Most of the literature on accretion onto this kind of objects refers to boson stars, although there exist at least one work \edit{simulating accretion onto Kerr superspinars (Section \ref{sec:superspinars}) and another one considering the case of wormholes (Section \ref{sec:wormholes})}.
Boson stars can be considered as surfaceless objects
when the scalar field that constitutes them
has very weak couplings with the particles of the
standard model. In such case, the exotic object interacts with the accretion flow only through gravity,
behaving as a clump of dark matter.
A similar reasoning can be applied to other \acp{ECO}
such as Proca stars; however, the available
literature is limited to the simulation of
accretion onto boson stars, likely due to the
simplicity of their construction.

\subsubsection{Accretion-Ejection in boson stars}
\label{sec:accretion-ejection-BS}
The earliest simulations are those of
\cite{Meliani2016}, which were performed
in pure hydrodynamics using the code \texttt{GR-AMRVAC}.
In such work, Meliani and collaborators explore different
accretion scenarios onto boson star metrics generated by
the spectral solver \texttt{KADATH} \cite{grandclement_kadath_2010}
for the case of a free scalar field (no self-interaction),
that is, subject to the potential
\begin{equation}
	\label{eq:quadratic-potential}
	U(|\Phi|) = m^2 |\Phi|^2\,,
\end{equation}
where $m$ is the mass of the scalar field.
Part of the aim of such work is to introduce tests to evaluate the
reliability of \ac{GRHD} codes in boson star spacetimes.

They start by considering spherical accretion onto boson stars,
the analogue of Bondi-Michel accretion onto \acp{BH} \cite{bondi_spherically_1952,Michel1972}, which is widely
used as a test for \ac{GRHD} codes.
These simulations are one-dimensional, in spherical symmetry, and exclude the origin,
placing the inner boundary at $r=1.8\ M$ or $r=3.1\ M$, where from now on $M$
is the ADM-mass of the boson star.
This avoids the fluid shocking with itself and accumulating at the center
of the boson star, preserving the stationary character of the Bondi-Michel solution.
The authors show that at small distances from the center of the boson star, the
behavior of matter differs considerably from the case of \acp{BH}.

The second test they present is a stationary torus in hydrodynamic equilibrium
around an axially symmetric boson star ($k=1$, $\omega=0.81\ m/ \hslash $).
The torus has constant specific angular momentum $\ell=4.206$ and is constructed in the way
of \cite{Abramowicz1978}.
The code evolves this stationary solution up to $t=10^3\ M$ showing still an
excellent agreement with the initial condition.

\begin{figure}
	\centering
	\includegraphics[width=0.5\linewidth]{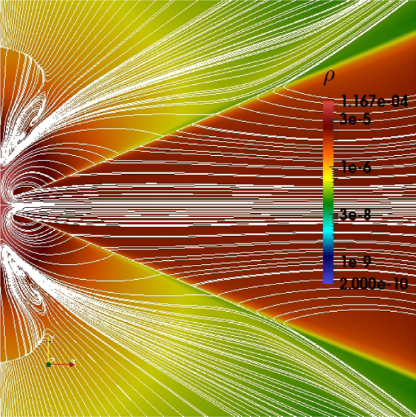}
	\caption{Accretion-ejection region near the center of the boson star in the last simulation by \cite{Meliani2016}. The color represents logarithmic density, and the white curves are velocity streamlines. From \cite{Meliani2016}.}
	\label{fig:inflow-outflow}
\end{figure}

Finally, they consider a non-spherical accretion scenario, where matter is
injected in a $20^\circ$ wedge around the equatorial plane.
This wedge is initially a Bondi-Michel flow truncated at $r=2\ M$.
The boson star used for this simulation is spherically symmetric ($k=0$, $\omega=0.771\ m/\hslash$),
and the simulation is \edit{performed in axisymmetry (2.5D)}.
As the system evolves, at first the inflow is decelerated by the increasing thermal pressure.
Soon, the the flow reaches the center of the boson star and produces a shock that propagates outwards
until it is halted by the supersonic inflow.
Later, a spherical structure sustained by pressure gradients appear and propagates out of the boson star.
The only direction in which matter can escape is along the poles. Figure \ref{fig:inflow-outflow}
shows the accretion-ejection region near the center of the boson star.
Part of the flow is redirected toward the accretion flow, while another fraction escapes
as a thermal wind.
The strong radial pressure gradient causes radial of the acceleration of the outflow,
which becomes supersonic for $r > 4\ M$.
Finally, interactions between inflow and outflow render the structure unstable,
and the shock front changes shape from spherical to triangular.
At later stages, the system reaches a quasi-steady state with some variability in the wind intensity.
The authors mention that this model does not indicate the presence
of Type I (i.e. thermonuclear) X-ray bursts, which are expected to be produced in some
models of accretion onto \acp{ECO}, and whose absence is considered to strengthen
the case of \ac{BH} candidates in X-ray binaries as true \acp{BH} \cite{yuan_constraining_2004}.
Overall, this simulation presents a scenario with a phenomenology that is
radically different from the case of accretion onto \acp{BH}, and
invites to consider more diverse scenarios for accretion-ejection flows in compact objects.


\subsubsection{Tidal disruption of clouds by boson stars}
\label{sec:tidal-disruption}
\ac{GRHD} simulations have also been employed to study the tidal disruption
of gas clouds by boson stars.
These studies were in part motivated by the pericenter passage in 2014 of the object G2,
likely a dusty, ionized gas cloud, in its eccentric orbit around \sgrastar \cite{gillessen_gas_2012}.
simulations of this event were performed by \cite{anninos_three-dimensional_2012} and
\cite{schartmann_simulations_2012}, among others.
However, for horizonless and surfaceless compact objects such as boson stars,
tidal disruption would produce a qualitatively different behavior as for \acp{BH}.
For example, boson stars allow closed orbits with very low angular momentum (including zero).
In contrast, for \acp{BH}, a cloud approaching with an angular momentum below some threshold
would lose a large fraction of its matter behind the event horizon.
In addition, matter that falls inside a boson star can in principle form structures inside and
continue to emit electromagnetic radiation that may reach a distant observer.

\begin{figure}
	\centering
	\includegraphics[width=0.9\linewidth]{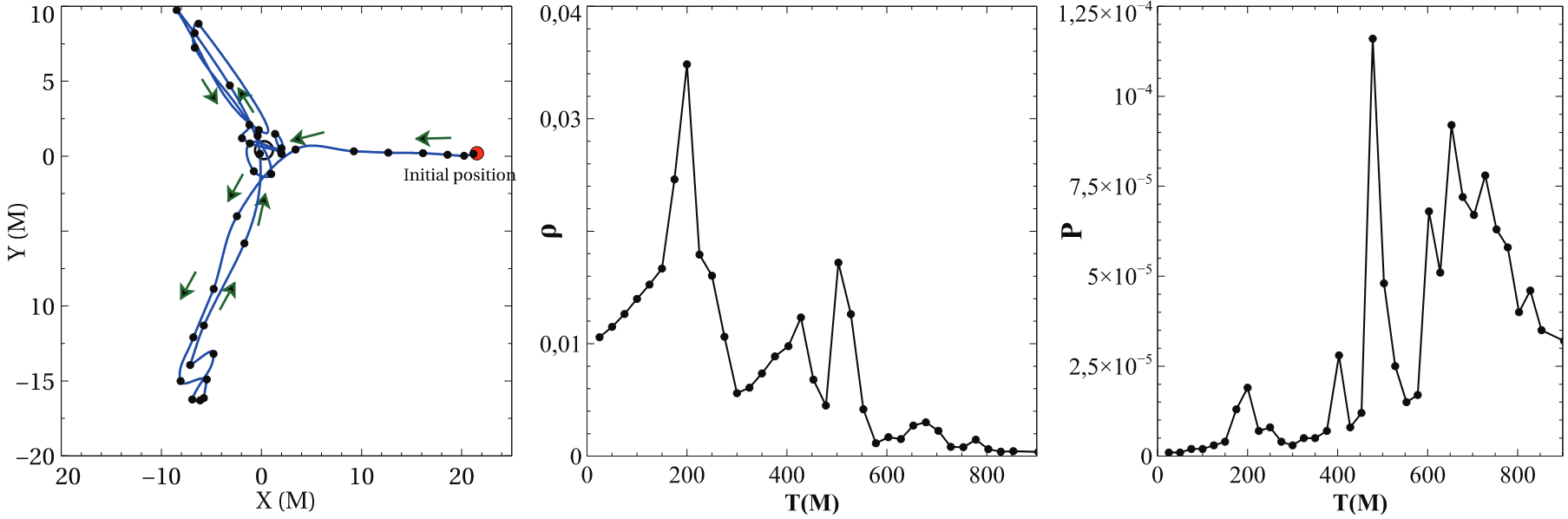}
	\caption{{\it Left panel:} Trajectory of the center of mass of a cloud initialized at $23\ M$
		away from a boson star with $k=1$ and $\omega = 0.8\ m/\hslash$.
		{\it Center and right panels:} Evolution of the density and pressure maxima
		for the same simulation, showing spikes that can be associated to flares.
		  From \cite{meliani_tidal_2017}.}
	\label{fig:tidal-disruption}
\end{figure}

The first \ac{GRHD} simulations tidal disruption of clouds by boson stars
appeared in \cite{meliani_tidal_2017}, also using the code \texttt{GR-AMRVAC}.
They consider three cases of falling clouds and two cases of boson stars.
Both of the boson stars are made from a free field and are rotating 
($k=1$, $\omega = 0.8\ m/\hslash$ and $k=4$, $\omega = 0.8\ m/\hslash$).
The gas cloud is initialized at rest at different distances from the boson star (23 and 3 $M$).
It has a Gaussian density profile and is in pressure equilibrium with the ambient medium.
Self-gravity is ignored, which is a reasonable approximation, as the gravitational force of the
compact object is expected to be much larger due to its proximity.
Simulations were two-dimensional and performed in Cartesian quasi-isotropic coordinates
on the equatorial plane of the boson stars.
This choice avoids spurious effects due to the singularity of the origin of spherical coordinates.

The trajectory of the center of mass of the clouds approximately follows a geodesic.
However, close to the boson star, pressure gradients caused by deformations and heating of the
cloud produce significant deviations from geodesic motion.
At this stage, the cloud acquires a complex structure, which can include shocks
propagating against the rotation of the star.

For the case of one of the clouds initialized at a large distance, the combination of
tidal stretching, frame dragging and deviations produced by hydrodynamic effects
cause the center of mass of
the cloud to scatter in the {\it opposite} direction with respect to the star rotation.
In general, the trajectories of the clouds are petal-shaped, with those initialized
at large distance showing more elongated petals (left panel of Figure \ref{fig:tidal-disruption}).

Following close encounters between the cloud and the boson star, a fraction
of the gas is teared off from the cloud by tidal forces.
The simulations by \cite{meliani_tidal_2017} show that most of this gas is recaptured
by the boson star and forms a disk at its interior, between the last stable orbit
and the center of the scalar field torus.

As it could be expected, the passage through the boson star is more violent for the
clouds that are initialized at a larger distance.
For those cases, the evolution of the pressure and density maximum exhibit
several spikes that could be observed as flares (center and right panels of Figure \ref{fig:tidal-disruption}).
The time between flares $\Delta T$
decreases significantly following the decreasing orbit size of the cloud
($\Delta T \sim 150,\ 250,\ 100,\ 60\ M$).
In contrast,  the cloud that is initialized close to the boson star requires
many passages to turn the gas expelled from the cloud into a disk.
The spikes in this case appear in larger number and at a more regular separation
in time ($\Delta T \sim 50 M$). In fact, the orbit of the center of mass
is almost completely well described by an ellipse affected by strong precession
with an important contribution from frame dragging.
The authors suggest that this simulation can be used to study the dynamics of
a hot spot orbiting a compact object.

\begin{figure}
	\centering
	\includegraphics[width=\linewidth]{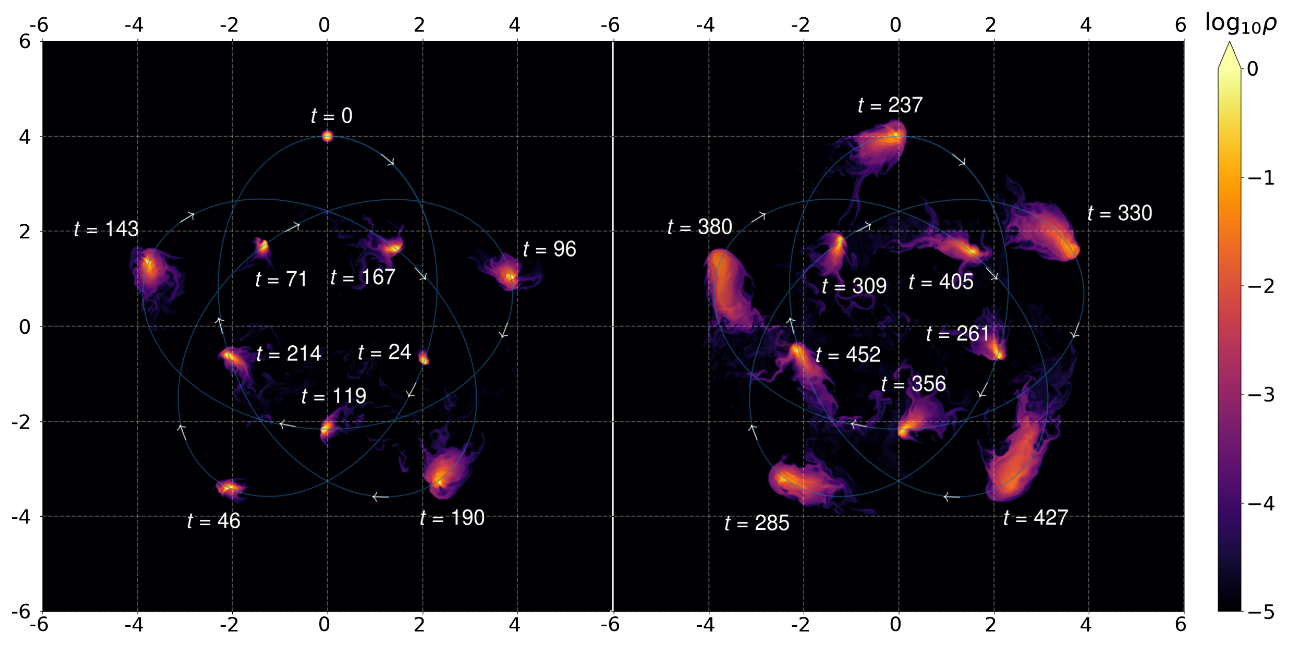}
	\caption{Composite image of a gas cloud closely following a five-petals geodesic around a solitonic star. The cloud experiences several compression-expansion cycles.
	From \cite{teodoro_tidal_2021}.}
	\label{fig:cloud-geodesic}
\end{figure}

Further simulations of tidal disruption by boson stars were presented by
\cite{teodoro_tidal_2021}.
In such work, the authors perform several simulations for the same
boson star model. This is no longer a mini-boson star, but a
solitonic boson star, subject to a sextic self-interaction potential
\begin{equation}
	\label{eq:solitonic-potential}
	U(|\Phi|)= |\Phi|^2 (m^2 + a |\Phi|^2 + b |\Phi|^4)\,.
\end{equation}
The parameters chosen for the scalar field are $m=1$, $a=-2$, and $b=1$.
The boson star considered has a frequency
$\omega = 0.27\ m/\hslash$ and is nonrotating. 
This object possess a compactness\footnote{\edit{The compactness of an
	object is defined as $\mathcal{C}=R/M$, where $R$ is the radius
	of the object and $M$ is its mass. Since bosonic stars are extended objects with no clear surface, the authors of \cite{teodoro_tidal_2021}
    define $R$ as the average radial coordinate weighted by the
    number density of the bosonic particles.}}
 $\mathcal{C} = 0.4$,
very close to the value $\mathcal{C} = 0.5$ characteristic of \acp{BH}.
The boson star considered in this work is nonrotating.
These simulations use the \ac{GRMHD} code \texttt{BHAC} \cite{Porth2016,Olivares2019a}
and similarly to those of \cite{Meliani2016,meliani_tidal_2017}
are performed in a Cartesian coordinate systems.
Also in this case, the cloud is not magnetized.

The authors explore four different scenarios for the encounter between a spherical cloud
and the boson star, namely, a zero angular momentum orbit, an elliptic orbit,
and two circular orbits for different sizes and initial positions of the cloud.
For the first three simulations, the cloud is very compact, with a size of $0.03\ M$.
Its initial position is at $r=4\ M$, that is, comparable with one initialized at $r=3\ M$ by \cite{meliani_tidal_2017}.
However, the observed behavior differs in important ways from that reported there.
This is due to the fact that the boson star is more compact and no longer rotating.

For the the zero-angular momentum case, the motion is now dominated by
local tidal forces, instead of the effect of frame dragging.
This results in a much faster deformation of the cloud
than for the rotating case, due to effects such as compression/expansion
cycles and collisions between the cloud and its tail or other debris.
For this reason, the cloud survives for a longer time for the case of the elliptic orbit,
and even longer for the case of a circular orbit, where interactions between
the cloud and its tail are more difficult.

Similarly to the $r=3\ M$ case reported by \cite{meliani_tidal_2017}, the
motion of the center of mass for the case with an elliptic orbit can be well described
by a geodesic. This is still a precessing ellipse (Figure \ref{fig:cloud-geodesic}),
although frame dragging does not contribute to this precession.

For the last boson star-cloud simulation, they place the cloud at a larger distance
of $r=10\ M$ and use a more extended cloud of size $0.3\ M$.
The initial velocity of the cloud is that \edit{corresponding to} a local circular geodesic.
They complement this simulation setup with an identical one where the boson star
is replaced by a Schwarzschild \ac{BH} in order to perform a comparison.
For this test, they calculate relevant diagnostics such as
the mass accretion rate, a proxy from Bremsstrahlung luminosity to produce a synthetic
lightcurve, and a shock detector.

They report that after the tearing of the cloud by tidal forces,
the debris starts to accumulate and form a ringlike structure around the boson star.
The structure also exhibits an extended spiral shock.
The persistence of this structure is in stark contrast with the \ac{BH} case,
where matter disappears behind the horizon, and could result in observable differences.
In fact, the synthetic lightcurve produced by the authors not only
raises orders of magnitude above that of the \ac{BH} simulation, but also persists
with a nearly constant luminosity for the rest of the simulation at $t = 1000\ M$.
In contrast the lightcurve for the \ac{BH} case drops to zero once
the entire cloud disappears behind the horizon, at $t=500\ M$.

\subsubsection{Magnetized accretion onto boson stars}
\label{sec:magnetized-accretion-BS}
Another set of observations that motivated studies on accretion onto boson
stars were the \ac{VLBI} campaigns to observe \sgrastar and M 87$^*$
by the \ac{EHT} Collaboration.
There were great expectations towards the first images of
\ac{SMBH} candidates at a resolution comparable to the angular size of the \acp{BH} shadows, and naturally towards the possibility of testing their \ac{BH} nature (see e.g. Ref. \cite{vincent_imaging_2016}).

To aid with the interpretation of observations, the \ac{EHT} Collaboration
created a library of \ac{GRMHD} simulations of Kerr \acp{BH} with several
spins and accreting in different modes (\ac{SANE} and \ac{MAD}).
These were later postprocessed by \ac{GRRT} codes to produce another library,
this time of synthetic images and spectra,
that could be compared with the real data \cite{EHT_M87_PaperV,EHT_SgrA_PaperV}.

In this context, the work of \cite{olivares_how_2020} applied the same
pipeline to study the observational properties of a boson star in the same accretion
regime as the \ac{EHT} targets, with the goal of assessing the possibility of
distinguishing it from a \ac{BH} by means of the \ac{EHT} observations.
To this end, the authors performed simulations of magnetized accretion
onto two members of the family of nonrotating mini boson stars, namely
$\omega M = 0.32$ (hereafter `model A') and $\omega M = 0.54$ (hereafter `model B').
Model A belongs to the unstable branch of the mini boson star family;
however, the choice of these models was not based on their
stability properties, but with the purpose of exploring two
qualitatively different behaviors that can take place also in more
general surfaceless and horizonless compact objects, and that
may play an important role in their observational properties,
as we will discuss later.

The \ac{GRMHD} simulation setup was similar to that used for the simulations
in the \ac{EHT} library. That is, a torus in hydrodynamic equilibrium
around a \ac{BH} was seeded with a poloidal magnetic field loop,
and was slightly perturbed by white noise added to the pressure.
This triggered the \ac{MRI}, which amplified the magnetic field,
and started transporting angular momentum outwards, causing the torus material
to accrete onto the central object.
Due to the prescription chosen for the magnetic field,
the simulations were in the \ac{SANE} regime, that is,
Maxwell stresses played a secondary role, and most of the angular momentum
transport was mediated by the \ac{MRI}.

The first and most important difference between the two boson star models originates
precisely at this stage. In order for the \ac{MRI} to be active,
the angular velocity profile of the accretion disk $\Omega(r) := u^r/u^t$
must decrease with the distance from the center of rotation, that is,
$d\Omega/dr < 0$, where $u^\mu$ is the 4-velocity of the fluid elements.
As the angular momentum is redistributed by the \ac{MRI},
$\Omega(r)$ acquires a profile close to that of circular timelike geodesics
$\Omega_K(r)$, as it is common for \ac{SANE} simulations
(even though not as close as, for example, in thin disk theory).
In the Kerr spacetime, $d\Omega_K/dr < 0$ for all radii, and the instability
is always active.
However, while this is also the case for boson star B,
for boson star A there is a radius $r_{\rm turn}>0$ such that $d\Omega_K/dr > 0$
for $r<r_{\rm turn}$. As $\Omega(r)$ gets closer to $\Omega_K(r)$,
it acquires a maximum where $d\Omega(r)/dr\approx 0$,
and the instability is suppressed.
With no other means of efficient angular momentum transport,
the accretion flow becomes stalled and forms a torus inside the boson star,
with its inner edge close to $r=r_{\rm turn}$.
This structure appears to be stable and lasts until the end of the simulation.

\begin{figure}
	\centering
	\includegraphics[width=0.7\linewidth]{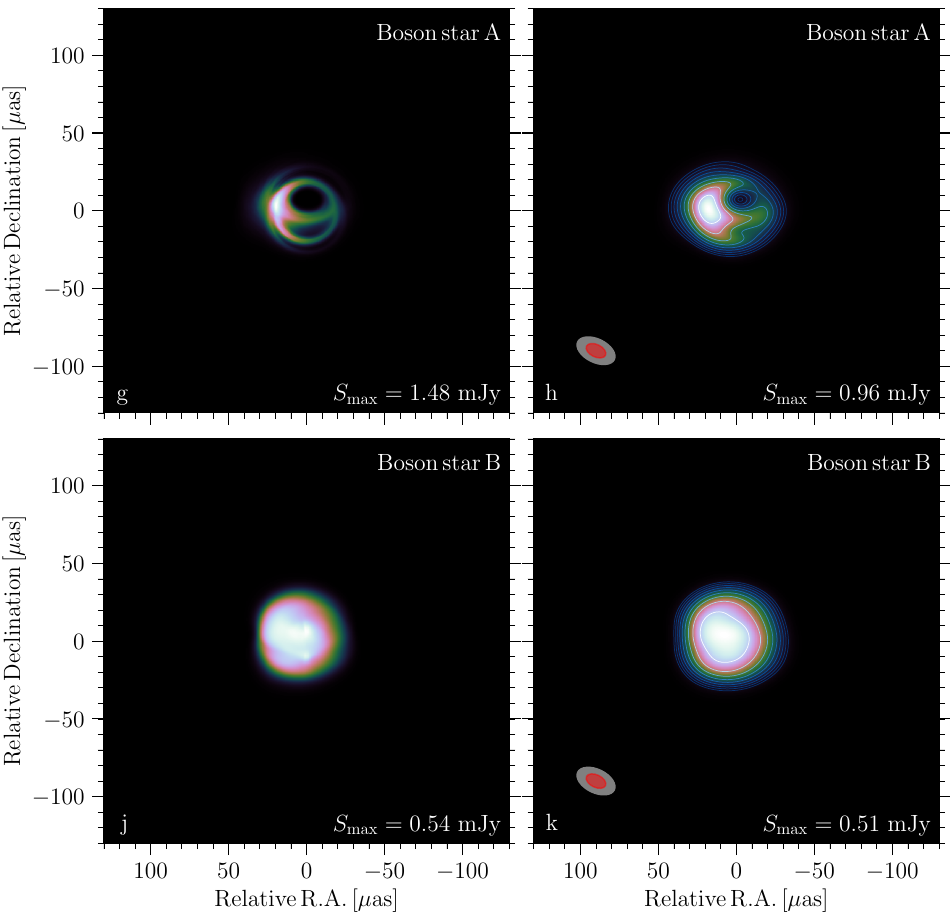}
	\caption{Ray-traced ({\it left column}) and synthetic ({\it right column})
		images at 230 GHz and inclination angle
		$60^\circ$ of boson star models A ({\it top row}) and
		B ({\it bottom row}).
		From \cite{olivares_how_2020}.}
	\label{fig:BosonStars-GRRT}
\end{figure}

When ray-tracing these simulations with the \ac{GRRT} code \texttt{BHOSS}
\cite{younsi_general_2012,younsi_modelling_2020},
the images produced for each of the boson star models are clearly different
(left column of Figure \ref{fig:BosonStars-GRRT}).
Boson star model B has the appearance of a spherical star-like object,
while boson star model A exhibits a central brightness depression,
resembling the typical image of a \ac{BH} shadow.
This brightness depression corresponds, however,
to the evacuated region at the center of the torus enhanced by gravitational lensing.
The differences are still clearly noticeable when convolving
the \ac{GRRT} image with 50\% of the \ac{EHT} beam in order to simulate
a resolution similar to that achievable by the instrument
(right column of Figure \ref{fig:BosonStars-GRRT}).
In particular, the central brightness depression survives.

However, one needs to be cautious before concluding that
mini boson stars can mimic supermassive \acp{BH}
in \ac{VLBI} observations.
One needs to consider, first, that the size of the brightness depression
for this boson star is considerably smaller than the one expected for
a Kerr \ac{BH} of the same mass (by 40\%), and second that
model A belongs to the unstable branch of the family.
As the instability can cause the star to collapse to a \ac{BH}
in a timescale of the order of less than an hour for \sgrastar and
less than a month for M 87$^*$, this particular model is clearly not viable.
It is therefore natural to ask whether there is a member of the family
that is stable and at the same time produces a brightness depression
of the appropriate size.

\begin{figure}
	\centering
	\includegraphics[width=0.7\linewidth]{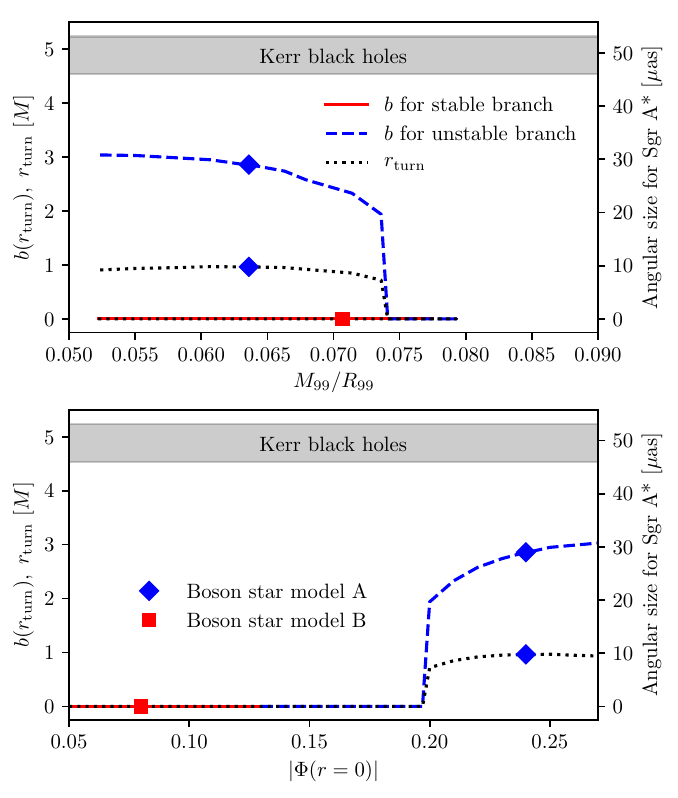}
	\caption{Radial position $r_{\rm turn}$ at which the \ac{MRI}
		is suppressed and apparent size of the central brightness
		depression after lensing $b(r_{\rm turn})$ for the family
		of mini-boson stars parametrized by compactness ({\it top panel})
		and central amplitude of the scalar field ({\it bottom panel}).
		The expected sizes of Kerr \ac{BH} shadows of the same mass
		are shown as a shaded region for comparison.
		From \cite{olivares_how_2020}.}
	\label{fig:CBD-boson}
\end{figure}

This can be estimated by calculating the radius $r_{\rm turn}$ and the size
of the depression after the enhancement by lensing, $b(r_{\rm turn})$,
both of which depend only on the metric and are easy to compute in
spherical symmetry, even when the
metric is numerical. The results for several members of the mini boson
star family are presented in Figure \ref{fig:CBD-boson}.
It can be seen that no members of the family produce
brightness depression with the required size to
be mistaken by Kerr \acp{BH} given the
\ac{EHT} resolution (of the order of 10 $\mu$as),
and that these are all produced on the unstable branch.
This numerical experiment therefore rules out
mini boson stars as viable models for
\sgrastar and M 87$^*$ under the accretion scenario
considered by \cite{olivares_how_2020} and under the assumption that
$\Omega(r) \approx \Omega_K(r)$.

It is important to remark that this result does not rule out
boson stars in general, and consequently neither other
horizonless compact objects, as viable models to \ac{SMBH} canditates.
Rather presents a scenario where they could act as \ac{BH}
mimickers by a previously unknown mechanism.
In fact, it has been shown that other horizonless compact
objects, namely Proca stars, could in principle produce
brightness depressions comparable to those of Kerr \acp{BH}
by this mechanism \cite{herdeiro_imitation_2021}.
On the other hand, it is conceivable that
other kinds of \acp{ECO} produce central brightness
depressions by different mechanisms,
such as the nonexistence of timelike circular geodesics
below some finite radius,
a property present in some rotating boson and Proca stars
\cite{sengo_imitation_2024}.

It is also important to remark that,
there are important differences that may help distinguishing it from
a \ac{BH} shadow. Most importantly, it would not be achromatic
phenomenon as it is the case for the latter.
For example, its size may vary with the observing frequency,
as it is dependent on the optical depth of the fluid.
Although at present the images released by the \ac{EHT}
Collaboration are at the single frequency of 230 GHz,
there are plans to provide the instrument with multi-frequency capabilities.
This would enable us to test the achromaticity of the \ac{BH}
shadow and potentially rule out scenarios as the one presented here.

Besides this, \cite{olivares_how_2020} report
other differences in simulations of magnetized accretion onto boson
stars with respect to what is observed in \ac{BH} simulations,
such as the absence of a jet, which can be explained in part
by the lack of rotation, and therefore of the absence of the
Blandford-Znajek mechanism;  and the presence of quasi-periodic
oscillations in the lightcurve of model B, which correspond
to the motion of the plasma density maximum around the center of the
boson star.

Before concluding this section, it is worth mentioning
that the results presented in \cite{olivares_how_2020}
have the limitation that they only consider
the \ac{SANE} accretion scenario.
However, increasing evidence favors
instead the \ac{MAD} scenario
for both \sgrastar and M87$^*$
\cite{EHT_M87_PaperV,EHT_SgrA_PaperV}.
It is uncertain whether the formation of the
central toroidal structure would also occur
for the latter, due to the larger
deviations from geodesic motion,
the secondary role played by the
\ac{MRI} in angular momentum transport, and the dynamical
importance of coherent magnetic fields, which may
introduce new unexpected effects.
This, however, highlights the importance of performing \ac{GRMHD}
for different models of \acp{ECO} under
a variety of accretion scenarios.

\subsubsection{\edit{Spherical accretion onto naked singularities}}
\label{sec:superspinars}
\edit{In Reference \cite{bambi_accretion_2009}, the authors performed
\ac{GRHD} simulations of quasi-spherical accretion onto a spacetime that
contains a naked singularity, namely Kerr spacetimes with dimensionless spin parameter $a\geq1$, also known as {\it superspinars} \cite{gimon_astrophysical_2009}.}

\edit{Their simulations adopt as initial and boundary condition a fluid with constant density and pressure, and excise the singularity from the domain by placing an
outflow boundary condition at $r=0.5\ M$ in Boyer-Lindquist coordinates.
This is justified by assuming that the singularity is a pathology of \ac{GR} that should be replaced by something else in the correct underlying theory, but that the detailed description of its unknown properties will not affect significantly the simulation results.
Simulations are 2.5D and are performed using the code PLUTO \cite{mignone_pluto_2007}.}

\edit{The authors found accretion patterns that differ qualitatively from the analogous case for \acp{BH}, namely Bondi-Michel accretion.
The properties of these accretion patters appear to depend on spin.
Below a critical value of $a\approx 1.4$, matter is unable to reach the central object due to short-distance repulsion, and instead forms a cloud that continues growing without reaching stationarity. In contrast, simulations above that critical value reach a stationary state in which the central object accretes through a thin equatorial stream. The authors discuss some of the astrophysical implications of both scenarios.
In particular, a harder and more intense radiation than for the case of accreting \acp{BH} is expected, which they link to the abundance of $\gamma$-rays observed in some blazars. They also discuss the possibility of violent events resulting from the collapse of the accreting system onto a black hole.}

\subsubsection{\edit{Magnetized accretion onto traversable wormholes}}
\label{sec:wormholes}
\edit{Another interesting example of accretion onto an horizonless, regular, exotic compact object is presented in Reference \cite{combi_general_2024}.
In that work, the authors simulate magnetized accretion onto traversable wormholes.
Specifically, the spacetimes they consider are described by the Simpson-Visser metric, a two-way traversable wormhole with no rotation.
This solution is similar to Schwarzschild from the outside, however it is not in vacuum and requires exotic matter to be sustained.
To perform the simulation, the authors assume that this exotic field does not interact with the accreting plasma, and that the latter does not back-react on the field.
The simulation set-up is similar to the one described for the boson star case in Section \ref{sec:magnetized-accretion-BS} and in typical \ac{BH}-torus simulations with \ac{MRI}-driven accretion. However, the disk is initialized only at one side of the wormhole.}

\edit{
The authors employ the code Harm3D and spherical Boyer-Lindquist type coordinates. More specifically, the wormhole metric is given by
\begin{equation}
	\label{eq:wormhole-metric}
	ds^2 = -f(r) dt^2 + f(r)^{-1} dr^2 + (r^2 + \ell^2)d\Omega^2
\end{equation}
where $r(r)\coloneqq (1 - 2M/\sqrt{r^2 + \ell^2})$ and, in contrast to typical spherical coordinates, $r\in(-\infty,+\infty)$.
The use of these coordinates allows to cover the whole spacetime with a single chart,
with the wormhole throat located at $r=0$.
In order to benefit from the properties of radially logarithmic grids, the radial coordinate above is mapped to a numerical coordinate $x$, such that
$r = c_r \sinh(x/c_r)$.
}

\edit{
The authors find that, after accretion starts, a portion of the plasma
settles at the throat, and forms a rotating cloud roughly $10^4$ times denser than the accretion disk. This cloud is hot and larger than the light ring, with its emission peaking on the far UV-X-rays for stellar-mass systems. Although no \ac{GRRT} calculations where performed, this would likely result in the absence of a central brightness depression that could mimic a \ac{BH} shadow.
}

\edit{
The system does not reach equilibrium. Instead, the cloud grows linearly with time, which could result in gravitational collapse onto a \ac{BH} in less than a Hubble time (also of stellar-mass systems).
}

\edit{
Another interesting finding in Reference \cite{combi_general_2024} is that matter falling into the wormhole powers a mildly relativistic thermal wind at the other side of the throat. The efficiency of this outflow (the energy output divided by mass accretion rate) appears to depend on the size of the throat, and for $\ell=2.1\ M$, this efficiency is of $\sim 20\%$. Astronomical observations from this other side of the wormhole would show a \ac{BH}-like system that exhibits outflows without an associated inflow, contradicting the typical phenomenology expected for \acp{BH}.
All of the effects described so far are consequences of the fact that the throat is a minimum of the gravitational potential. This feature is shared by most traversable wormholes, which indicates that these results could apply to many other configurations.
}

\edit{
The presence of the central cloud and possible absence of a central brightness depression contradicts some previous expectations on wormhole images, which highlights the importance of performing \ac{GRMHD} simulations in these spacetimes.
Other interesting question that could be answered by future simulations are the effects of spacetime rotation and different initial conditions for the accretion flow.
}

\subsection{Non-Kerr black holes}
\label{sec:Non-Kerr-BHs-GRMHD}

As noted above, the Kerr metric \cite{Kerr:1963ud} can be used to describe stationary and axisymmetric vacuum \ac{BH} spacetimes. It is obtained as a solution of the dynamical equations of \ac{GR} and also in several other alternative classical gravity theories, e.g., those that admit \ac{GR} as a smooth limit. 

Non-Kerr \acp{BH} have broadly been considered in two contexts. First, allowing for the presence of self-gravitating matter and solving for the appropriate gravitational as well as matter field equations yields a variety of non-Kerr metrics. Examples in \ac{GR} of such \textit{solution metrics} include (a) the Reissner-Nordstr{\"o}m (RN) spacetime (see, e.g., \cite{Misner+1973}), which contains a static electromagnetic (EM) field, (b) the Kerr-Newman (KN) spacetime \cite{Newman+1965b}, which contains a stationary EM field, (c) the Hayward spacetime \cite{Hayward2006}, which contains a static anisotropic fluid (with no viscosity or heat fluxes), (d) the \ac{GMGHS}; \cite{Gibbons+1988, Garfinkle+1991} spacetime, which contains a dilaton scalar field as well as an EM field, and (e) the Kerr-Sen (KS) spacetime \cite{Sen1992}, which contains a dilaton, an EM, and an axion field. The nonspinning RN \ac{BH} and its spinning generalization, the KN \ac{BH}, both admit two horizons and also possess curvature singularities at $r=0$ (in appropriate coordinates). The Hayward \ac{BH} also admits two horizons but is regular at the center, $r=0$. The \ac{GMGHS} \ac{BH} has a single horizon and the same surface gravity as a Schwarzschild \ac{BH}, independent of the strength of its field content. Together with the KS \ac{BH}, its spinning counterpart, these arise as solutions to the low-energy effective limit of the heterotic string theory. Examples of non-Kerr \acp{BH} in alternative theories include the aforementioned Einstein-scalar-Gauss-Bonnet \acp{BH}. 

Perhaps the key ambition behind ascertaining whether or not non-Kerr \ac{BH} solutions can be distinguished from Kerr \acp{BH} is to probe both the various types of \ac{BH} models as well as their underlying theories of gravity and fields for consistency with observations performed on horizon scales.

The second context that has generated interest in the study of non-Kerr \ac{BH} spacetimes is the attempt to gather further observational evidence for the Kerr hypothesis. This is achieved by working with \textit{parametrized metrics}, which agnostically deform away from the Kerr metric smoothly, making no reference to the field equations. Several parametrized metrics have been designed to capture qualitatively different aspects of possible metric deviations, e.g., the Johannsen-Psaltis (JP) metric \cite{Johannsen+2011, Johannsen2013}, the modified gravity bumpy Kerr metric \cite{Vigeland+2011, Gair+2011}, Konoplya-Rezzolla-Zhidenko metric \cite{Konoplya+2016}, etc. 

The Event Horizon Telescope (EHT) observations of M87$^*$ and Sgr A$^*$ \cite{EHTC+2019a, EHTC+2022a} are particularly suited to such ``tests'' of the space of stationary non-Kerr metrics \cite{Johannsen+2010, Mizuno2018, EHTC+2019f, Psaltis+2020, Kocherlakota+2021, EHTC+2022f}. Since the two EHT sources are supermassive ultracompact objects that are accreting at extremely low rates ($\sim 10^{-3}M_\odot\mathrm{yr}^{-1}$ and $\lesssim 10^{-7} M_\odot\mathrm{yr}^{-1}$ respectively; \cite{EHTC+2019a, EHTC+2022a}), their spacetimes are stationary to a very good approximation.


\subsubsection{Hairy black hole moving through an uniform medium}
\label{sec:BHL-hairy}
As mentioned previously, an important motivation to explore
the consequences of the existence of ultralight bosonic fields
comes from dark matter.
To understand the effects of the scalar field in a very general
situation,
the authors of \cite{cruz-osorio_bondi-hoyle-lyttleton_2023}
simulated the flow pattern around a \ac{BH} with synchronized
scalar hair, from now on \ac{HBH}, moving through a uniform medium.
For a star or a \ac{BH}, this scenario is known as 
\ac{BHL} accretion \cite{hoyle_effect_1939,bondi_mechanism_1944}.
\ac{BHL} accretion onto a \ac{BH} is relevant
to model several situations of interest to Astronomy,
such as a wandering \ac{BH} moving through the interstellar medium,
a \ac{BH} accreting low angular momentum winds
from a binary companion, or a \ac{BH} in the so called
common envelop phase of binary evolution.

The \ac{HBH} model employed is that from \cite{Herdeiro:2014goa},
where the scalar field is subject to a quadratic potential
(equation \ref{eq:quadratic-potential}).
As it is customary, simulations are performed in the rest frame
of the \ac{BH}. In this frame, the fluid is moving at Mach number
$\mathcal{M} = v_\infty/c_{s,\infty} = 5$, where
$v_\infty$ and $c_{s,\infty}$ are the asymptotic fluid velocity
and sound speed, respectively.
Four simulations are performed, characterized by different
ratios between the mass of the scalar cloud and the ADM mass
($p:=M_\Phi/M_{\rm ADM}$),
and between the angular momentum of the scalar cloud
and the total ADM angular momentum ($q:=J_\Phi/J_{\rm ADM}$).
The specific values for all simulations were:
$p=$~(0.53, 0.749, 0.909, 0.982) and $q=$~(0.128, 0.846, 0.997, 0.998), in the same order..
The simulations are carried out in pure \ac{GRHD} and in two dimensions on the equatorial plane of the \acp{BH}.
They employ spherical coordinates and are performed using
the code \texttt{BHAC}.

The simulations reveal a flow pattern that is qualitatively
similar to that of \acp{BH} in \ac{BHL} accretion,
characterized by an upstream bow shock and downstream shock cone.
However, the quantitative properties such as the opening
angle of the shocks, the location of the stagnation points
in units of the \ac{BH} radius, and the mass accretion
rate, depend strongly on the relative
importance of the scalar field over the central \ac{BH}.
The authors fit analytical expressions for the mass accretion
rate to the numerical results. These could be used to study
\ac{BH} formation and growth scenarios in the presence of
ultralight scalar fields.

Similarly to the case of \ac{BHL} accretion onto
\acp{BH}~\cite{donmez_development_2011},
the authors observe \acp{QPO} as a result of pressure
modes trapped inside the shock cone.
The authors study these oscillations by placing detectors
close to the stagnation point and producing time series
of rest-mass density.
They calculate power spectral densities from these time series,
and study the effect of the presence of the scalar cloud.
Finally, they speculate about the possibility of linking these
\acp{QPO} with those observed in the X-ray lightcurves
of \sgrastar and some microquasars,
and compare them with the observational data.

\subsubsection{Thin disk accretion onto non-Kerr parameterized black hole}
\label{sec:parameterized-BHs-GRMHD}

The work of \cite{Nampalliwar2022} 
performs \ac{GRMHD} simulations of
accretion onto non-Kerr \acp{BH}.
However, the spacetimes considered
do not belong to any particular
theory or model.
They come instead from the
theory agnostic parameterization
known as the Johannsen metric~\cite{johannsen_regular_2013}.
In this parameterization,
four radial functions quantify deviations
from the Kerr metric, which
are written as series in powers
of $M/r$.
The parameterized metric retains
the separability of the Kerr metric
that leads to the Carter constant.

Theory-agnostic approaches
have the advantage that the metric can
be easily written in a convenient form,
and can provide a handle on the effect
of deviations from the Kerr metric in tests
where it is taken as the null hypothesis.
The motivation for \cite{Nampalliwar2022}
comes precisely from the possibility
of testing the Kerr hypothesis
for stellar mass \acp{BH} and \acp{SMBH}
using X-ray spectroscopy and breaking
degeneracies in the ISCO size between
the \ac{BH} spin and 
deviations from Kerr geometry.

For the \ac{GRMHD} simulations, the authors
employ a horizon-penetrating version of
the Johannsen metric. They also keep
only the leading terms of the expansion of
the deviation functions, which results in
four deviation parameters.
Simulations are performed using the
code \texttt{HARMPI} \cite{HARMPI2019},
which is based on the \texttt{HARM} family~\cite{gammie_harm_2003}.
The authors report seven simulations,
most of which are performed
in \edit{2.5} dimensions on the meridional plane.
Only one simulation is performed in
three dimensions as a check for the
validity of the two-dimensional results.
The initial conditions for the simulations
are the similar to those 
described in Section \ref{sec:magnetized-accretion-BS}. However,
since X-ray reflection spectroscopy
is mainly focused on thin accretion disks,
a cooling function is added in order to
keep the disk thin (with aspect ratio
$H/R=0.12$).
Several of the spacetime metrics used share
the same value of the ISCO, even when
they have different non-Kerr parameteres
switched on and off, and some of them
are purely Kerr cases.

The simulation results show that
the accretion pattern is similar to that
seen in typical \ac{GRMHD} simulations
in the Kerr spacetime. The authors also
conclude that semianalytic thin disk models
for reflection spectra are a reasonable approximation
to their simulations.
Finally, they show that even though the total
flux can be similar between Kerr and non-Kerr
cases, iron line plots
are visible different, strengthening the case for
X-ray spectroscopy as a sensitive test for deviations with respect to Kerr metric.

To finalize this Section, we would like to
remark the usefulness of theory-agnostic parameterized
metrics for parameter space surveys that may allow
to constrain theories of gravity.
Even though currently it is impractical
to perform such surveys on \ac{GRMHD} simulations,
due to their high computational cost,
understanding the effect of changing the parameters on
\ac{GRMHD} simulations can be used
to inform semi-analytic models (such as thin disks)
for which surveying the parameter space can be made in
a much more efficient way.


\subsubsection{Magnetized accretion onto dilaton-axion \acp{BH}: approaching naked singularities}
\label{sec:axion-dilaton-GRMHD}

As noted in Section \ref{sec:magnetized-accretion-BS}, the \ac{EHT} observations have been used to produce images of its two main targets, which have been compared against synthetic images produced from state-of-the-art numerical simulations. While such simulations have mostly been performed using the Kerr metric, the first simulations of non-Kerr spacetimes were reported in Ref. \cite{Mizuno2018}.

\begin{figure}
\centering
\includegraphics[width=\columnwidth]{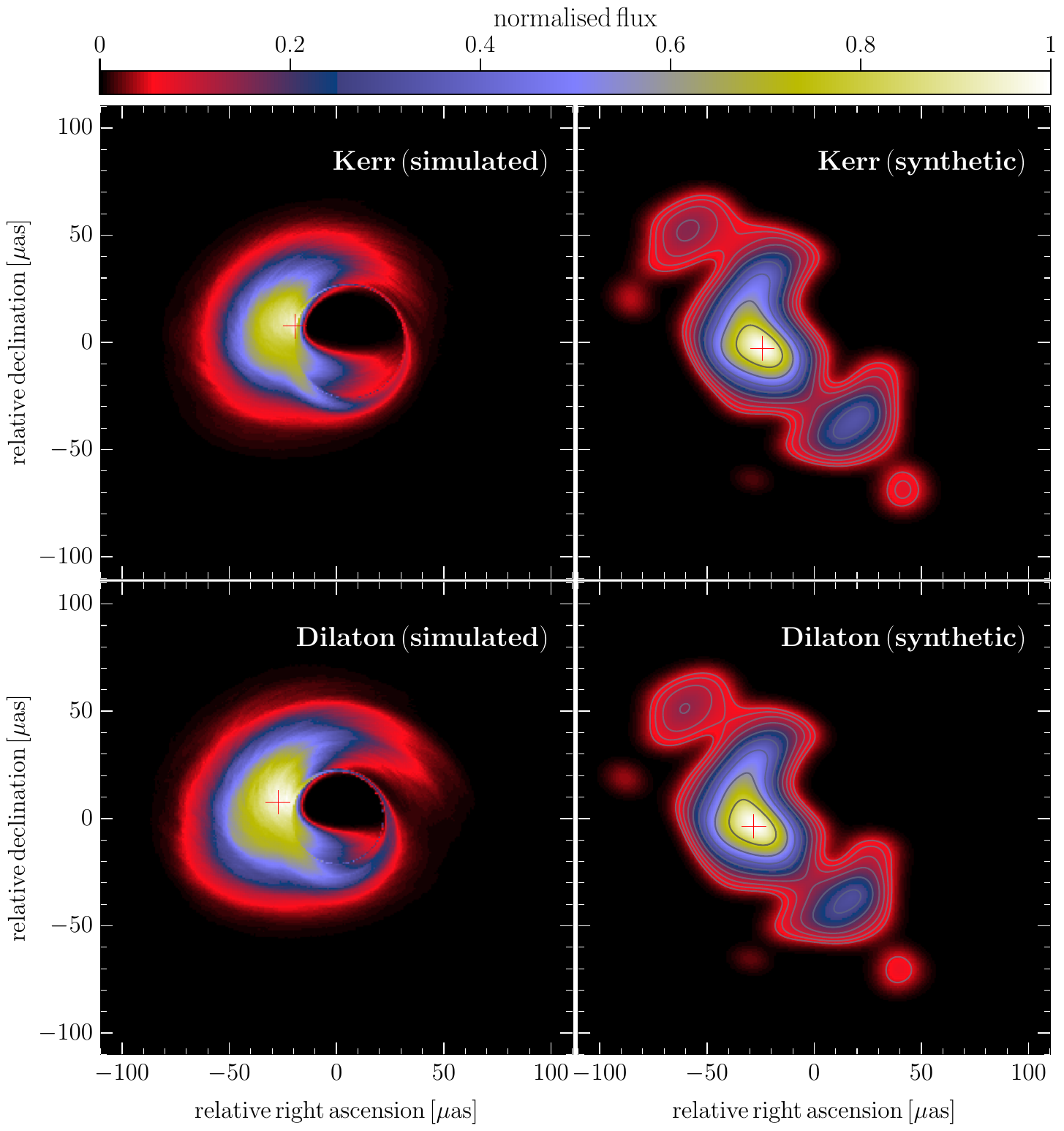}
\caption{Ray-traced ({\it left column}) and synthetic ({\it right columns}) images of a Kerr ({\it top row}) and
a dilaton ({\it bottom row}) \ac{BH} simulated in \cite{Mizuno2018}. The two \acp{BH} are matched by the \ac{ISCO}. Credit image: Christian Fromm, Yosuke Mizuno, Ziri Younsi.}
\label{fig:Fig1.7_Kerr_v_Dilaton_BH}
\end{figure}

The goal of such work was to assess the capabilities
of the \ac{EHT} array to distinguish between different theories of gravity by observing the \ac{BH} shadows.
To this end, three different simulations
of spherically symmetric non-Kerr solutions were performed.
These were later postprocessed via \ac{GRRT}
calculations to produce images, which where compared
to a simulated image of a Kerr \ac{BH}
(\edit{see Figure \ref{fig:Fig1.7_Kerr_v_Dilaton_BH}}).
These solutions came from the 
Einstein-Maxwell-Axion-Dilaton theory~\cite{garcia_class_1995},
which introduces two hypothetical scalar fields
(the axion and the dilaton).
The three cases considered were selected
by matching one of the relevant radii to the value of a Kerr
\ac{BH}, namely, the \ac{ISCO}, the photon ring
or the event horizon.

Fig. \ref{fig:Fig1.7_Kerr_v_Dilaton_BH}, taken from there, shows ray-traced and synthetic time-averaged images (for the \sgrastar
parameters)
of a Kerr \ac{BH} (top row) and a (non-rotating) GMGHS or ``dilaton'' \ac{BH} (bottom row), matched to the \ac{ISCO}.
The right column shows synthetic images taking into account
the finite resolution of the \ac{EHT} array, and demonstrates the difficulty in distinguishing the two \acp{BH} from a direct comparison of their images. The inclusion of
uncertainties from the instrument and observing conditions
renders the two \acp{BH} practically indistinguishable.

As a follow up of the work by \cite{Mizuno2018},
\cite{Roder:2023oqa} produced new simulations
in these nonspinning spacetimes.
These included the \ac{MAD} scenario
and more sophisticated radiation models.
Some interesting results included
a wider jet opening angle and a higher magnetization
in the case of the Kerr spacetime.

However, it was also observed that
the emission model has a larger effect on the images
as compared to the spacetime metric or the accretion
model, leading to the conclusion that
the imaging approach may not be sufficient
to distinguish the two spacetimes with the current
resolution of the \ac{EHT}, as of 2023.

Until recently, accretion in non-Kerr spacetimes has only been explored in nonspinning spacetimes. In Ref. \cite{Nampalliwar2022}, \edit{the first 2.5D} \ac{GRMHD} simulation in a spinning non-Kerr spacetime was performed, which we discussed in Sec. \ref{sec:parameterized-BHs-GRMHD} above. In Refs. \cite{Chatterjee+2023a, Chatterjee+2023b}, a large number ($\sim 100$) of 3D \ac{MAD} \ac{GRMHD} simulations in spinning non-Kerr spacetimes were performed. The first of these studies focussed on accretion onto Kerr-Sen (KS) or ``dilaton-axion'' \acp{BH}, extending the earlier work of Refs. \cite{Mizuno2018, Roder:2023oqa} discussed above. The metric for these spacetimes takes the form given in eq. \ref{eq:AA_Metric_BL} above with $A = r(r+2D)$ and $B=r^2 - 2(M-D)r$.

The right panels of Fig. 1 in Ref. \cite{Chatterjee+2023a} show the time-averaged 230 GHz (the EHT observing frequency) images of four different \acp{BH}, two nonspinning and two spinning ones. These are a Schwarzschild \ac{BH}, a Kerr \ac{BH} with dimensionless $a_*=0.5$, a near-extremal dilaton \ac{BH} with dilaton charge $D = 0.995M$, and a near-extremal dilaton-axion \ac{BH} with $D = 0.5M$ and $a = 0.49M$. Since the dilaton spacetime with $D=1M$ describes a \textit{naked singularity} spacetime, the dilaton \ac{BH} image provides hints regarding the image of such \textit{exotic} possibilities.

The M87$^*$ \ac{BH} parameters are chosen when constructing these images (mass, distance, inclination, etc.). The bright ring in each image is the photon ring \cite{Johnson+2020}, a collection of all higher-order images of the entire horizon-scale accretion flow. The photon ring is tightly tied to the \ac{BH} shadow boundary curve, which is related to its light rings. Observable differences between the sizes of these rings are clear to see in these synthetic images. These differences may be possible to distinguish with future space-based very long baseline interferometry observations, which hope to resolve the photon rings of M87$^*$. 

Similarly, images for three classes of Johannsen-Psaltis parametrized \ac{BH} spacetimes were constructed in Ref. \cite{Chatterjee+2023b}. The parameters allow one to deform smoothly away from a Kerr geometry, and control qualitatively different aspects of the spacetime. For example, the parameter $\alpha_{13}$ changes the size of the shadow boundary curve whereas the parameter $\alpha_{22}$ has a predominant effect on its shape (it changes the \ac{BH} quadrupole moment) \cite{Johannsen2013}. Thus, measuring the size and shape of the \ac{BH} shadow from observations can be used to impose constraints on such parameters, and build confidence in the Kerr hypothesis. 

\subsubsection{Energy Extraction from Non-Kerr Black Holes}
\label{sec:energy-extraction}

Thus far we have focussed on the importance of \ac{GRMHD} simulations in producing realistic synthetic images of accreting \acp{BH}, which can directly be compared with the highest angular-resolution images of accreting astrophysical \acp{BH}. Such simulations, however, provide yet another rich vein of physical insight: Spinning \acp{BH} that accrete magentized plasma generically produce ``jets,'' which are relativistic outflows of magentically-dominated plasmas. Indeed, astrophysical jets are frequently observed across the electromagnetic spectrum (see, e.g., the Hubble Space Telescope image of the M87 jet in Ref. \cite{Biretta+1999}), \edit{and much work has been devoted to explaining their production mechanism}.

\edit{One of the leading proposals is the \ac{BZ} mechanism \cite{Blandford+1977}, which is an electromagnetic Penrose process involving spinning \acp{BH}} \cite{Lasota+2014}. The jet outflow power generated from this mechanism is heuristically understood to be given as $P \approx a^2B^2r_{\mathrm{g}}^2c$, where the total energy density in the outflowing plasma is effectively given by the square of the magnetic field strength $\rho \sim B^2$, the outflow speed is $\approx c$, the area of cross-section of the outflow is roughly determined by the gravitational radius $r_{\mathrm{g}} = GM/c^2$, and the spin of the \ac{BH} is an important modulating factor. The above can be rewritten as $P \approx \Phi^2 (a/r_{\mathrm{g}}^2)$, where $\Phi = Br_{\mathrm{g}}^2$ is the magnetic flux on the horizon, and depends on the specifics of the accretion flow. For Kerr \acp{BH} with finite spin, it was shown in Ref. \cite{Tchekhovskoy+2010} that the remainder $(a/r_{\mathrm{g}})^2$ should be replaced by $\Omega_{\mathrm{H}}^2$, i.e., the square of the horizon angular velocity, which is determined purely by the spacetime geometry, i.e., $P \approx \Phi^2\Omega^2_{\mathrm{H}}$. The outflow efficiency $\eta$ defined as $\eta := P/\dot{M}c^2$ then measures the amount of outflow power per unit of rest mass energy accreted onto the \ac{BH}. If $\eta > 1$, a portion of the outflow energy must be extracted from the ``spin energy'' of the \ac{BH} itself, via an (electromagnetic) Penrose process.
\edit{Assuming the \ac{BZ} mechanism, the estimate of the jet power has already been used to test the Kerr hypothesis, as proposed in Ref. \cite{bambi_attempt_2012}. For further discussion, we direct the reader to an excellent review of this topic in Ref. \cite{Hawley+2015}}. 

\begin{figure}
\centering
\includegraphics[width=\columnwidth]{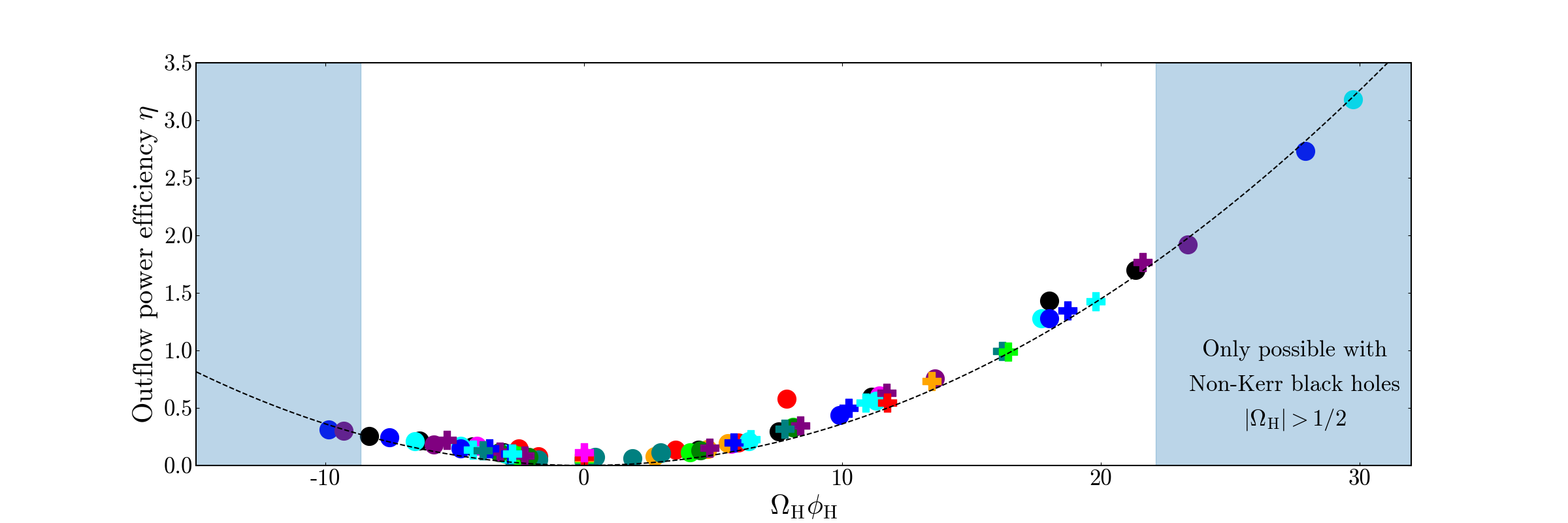}
\caption{Shown here is the outflow efficiency for Kerr-Sen \acp{BH} and Johannsen-Psaltis \acp{BH}, and their consistency the \ac{BZ} prediction. This demonstrates the universality of energy extraction via an electromagnetic Penrose process from various spinning \acp{BH}, i.e., Kerr as well as non-Kerr \acp{BH}. Adapted from Refs. \cite{Chatterjee+2023a, Chatterjee+2023b}.}
\label{fig:Fig1.10_Kerr_v_Sen_BHs}
\end{figure}

Fig. \ref{fig:Fig1.10_Kerr_v_Sen_BHs}, taken from Refs. \cite{Chatterjee+2023a, Chatterjee+2023b}, shows the maximum outflow efficiency $\eta$ computed from \ac{MAD} \ac{GRMHD} simulations in $\sim 100$ different Kerr and non-Kerr \ac{BH} spacetimes. To compare we also show in the dashed line the analytic prediction $\eta_{\mathrm{BZ}} \propto (\phi_{\mathrm{H}}\Omega_{\mathrm{H}})^2$, where $\phi_{\mathrm{H}} := \Phi/\sqrt{\dot{M}}$ is the (saturated) dimensionless horizon magnetic flux. This figure demonstrates how the \ac{BZ} prediction provides an excellent universal description of the jet power in \ac{GRMHD} simulations. \edit{Comparing these universal jet power predictions against actual observations may provide concrete evidence towards establishing the \ac{BZ} mechanism as the key driver of astrophysical jets in active galactic nuclei}. We also note that non-Kerr \acp{BH} can produce significantly higher jet powers than Kerr \acp{BH}. The application presented in this Section also showcases the insight that can be gained from libraries of simulations, that in this case were generated efficiently by the GPU-accelerated code \texttt{H-AMR}.

Finally, it is worth noting that, assuming the \ac{BZ} mechanism, estimates of the jet power can be used to test the Kerr hypothesis, as first proposed in Ref. \cite{Bambi+2012b}.


\section{Future perspectives}
\label{sec:conclusion}
In this Chapter, we have presented a landscape of some
of the most important families of \acp{ECO},
summarized some of the considerations that are in
order to performed \ac{GRMHD} simulations
on some of these objects, and have covered
the existent literature on the subject
while trying to provide the motivation behind each
work.

As it can be noticed, the field is extremely young,
and there are still many research directions to explore.
For instance, regarding horizonless compact objects,
it would be interesting to verify
via \ac{GRMHD} simulations whether it is really
possible to produce central brightness depressions
similar to \ac{BH} shadows for objects that
are dynamically stable, either by the mechanism
describe in Section \ref{sec:magnetized-accretion-BS},
or by different mechanisms.
Also, there are currently no simulations of rotating
horizonless compact objects, even though such solutions
exist. These could be used to study the possible
formation of jets, as well as to understand
what happens realistically in spacetimes that have
peculiar geodesic properties, such as the non-existence
of circular timelike orbits.
Currently there are also no simulations of \acp{ECO}
that possess surfaces. A model for the interaction
between the surface and the accretion flow could yield
new observational signatures to constrain the existence
of these objects.
A different problem, also related to boundary
conditions, is that of the treatment of naked singularities
within a fluid simulation.
This research line would allow to extend the
work on imaging such exotic objects,
which so far has considered only ray-traced images
with no underlying \ac{GRMHD} model.

Another natural extension of existing simulations
could come from taking into consideration
interactions between the additional fundamental fields
that are present in many \ac{ECO} models.
In fact, some of these fields (e.g. dilatons and axions)
are expected to interact with matter or the electromagnetic
field, but these interactions have so far been neglected.
Such simulations could help predicting signatures
of new fundamental fields even in cases where the latter
are not sufficiently strong to produce a significant
gravitational interaction.

Finally, another direction that we expect to be very fruitful
is the production of libraries of simulated
data. Besides allowing us to identify trends
as in the works described in Section
 \ref{sec:energy-extraction},
this would allow to systematically constrain
parameters for \acp{ECO} based on observations,
and ultimately test their existence.


\begin{acknowledgement}
We thank Z. Meliani, M. C. Teodoro, C. Fromm, Y. Mizuno and Z. Younsi for granting permission to reproduce their figures.
HO is supported by the Individual CEEC program - 5th edition funded by the Portuguese Foundation for Science and Technology (FCT).
This work is supported by the Center for Research and Development in Mathematics and Applications (CIDMA) through the
Portuguese Foundation for Science and Technology (FCT - Fundação para a Ciência e a Tecnologia), references UIDB/04106/2020, UIDP/04106/2020 (https://doi.org/10.54499/UIDB/04106/2020 and https://doi.org/10.54499/UIDP/04106/2020). The authors acknowledge support from the projects PTDC/FIS-AST/3041/2020, CERN/FIS-PAR/0024/2021 and 2022.04560.PTDC. This work has further been supported by the European Union’s Horizon 2020 research and innovation (RISE) programme H2020-MSCA-RISE-2017 Grant No. FunFiCO-777740 and by the European Horizon Europe staff exchange (SE) programme HORIZON-MSCA-2021-SE-01 Grant No. NewFunFiCO-10108625.
PK acknowledges support from grants from the Gordon and Betty Moore Foundation (GBMF-8273) and the John Templeton Foundation (\#62286) to the Black Hole Initiative at Harvard University.
\end{acknowledgement}

\bibliography{Refs-Exotic-Compact-Objects}


\end{document}